\documentclass[twocolumn]{article}

\usepackage{url}

\usepackage[numbers]{natbib}

\usepackage{amsmath} 
\usepackage{multirow} 
\usepackage{rotating}

\usepackage{authblk}

\begin{document}

\date{}

\title{Capability of the HAWC gamma-ray observatory for the indirect detection of ultra-high energy neutrinos}
\author[1]{Hermes Le\'on Vargas \thanks{Corresponding author: hleonvar@fisica.unam.mx}} 
\author[1]{Andr\'es Sandoval}
\author[1]{Ernesto Belmont}
\author[1]{Rub\'en Alfaro}

\affil[1]{Instituto de F\'isica, Universidad Nacional Aut\'onoma de M\'exico, Apartado Postal 20-364, M\'exico D.F. 01000, M\'exico.}

\renewcommand\Authands{ and }

\maketitle



\begin{abstract}
The detection of ultra-high energy neutrinos, with energies in the $PeV$ range or above, is a topic of great interest in modern astroparticle physics. The importance comes from the fact that these neutrinos point back to the most energetic particle accelerators in the Universe, and provide information about their underlying acceleration mechanisms. Atmospheric neutrinos are a background for these challenging measurements, but their rate is expected to be negligible above $\approx$ 1 $PeV$. In this work we describe the feasibility to study ultra-high energy neutrinos based on the Earth-skimming technique, by detecting the charged leptons produced in neutrino-nucleon interactions in a high mass target. We propose to detect the charged leptons, or their decay products, with the High Altitude Water Cherenkov (HAWC) observatory, and use as a large mass target for the neutrino interactions the Pico de Orizaba volcano, the highest mountain in Mexico. In this work we develop an estimate of the detection rate using a geometrical model to calculate the effective area of the observatory. Our results show that it may be feasible to perform measurements of the ultra-high energy neutrino flux from cosmic origin during the expected lifetime of the HAWC observatory.
\end{abstract}


\section{Introduction}
\label{intro}

The first evidence of ultra-high energy neutrinos (in the $PeV$ energy range) from extraterrestrial origin was recently reported \cite{ice-cube1}. This opened a new field in astroparticle physics that will allow the identification and characterization of the most powerful particle accelerators in the Universe. Neutrinos are not affected by the electromagnetic or strong interactions, and thus point back to the source where they were produced, unlike charged cosmic rays. Gamma-rays are another cosmic probe that provides information about the acceleration mechanisms that occur in astrophysical sources. Due to this, there are several dedicated instruments, both ground or space based, performing a continuous survey of the Universe characterizing gamma-ray sources, e.g. the Fermi Gamma-ray Space Telescope and the imaging atmospheric Cherenkov telescopes H.E.S.S., MAGIC, VERITAS and FACT.\\ 
The HAWC observatory was designed to detect and characterize the sources of high-energy gamma-rays in the energy range between 100 $GeV$ and 100 $TeV$ \cite{hawc-gamma-rays} and started full operations in April 2015. It is a ground based instrument that detects atmospheric showers by measuring with high precision the arrival time of the particles that compose the air showers. This is done via the Cherenkov light produced by the air shower particles as they enter the 300 water detector tanks that constitute the observatory, with a total water volume of 54 million $litres$. At energies above 100 $TeV$, gamma-rays suffer strong absorption from pair production with photons from the cosmic microwave background radiation that strongly reduces their mean free path \cite{opacity}. For this reason, ultra-high energy neutrinos may be a better tool to  study the most energetic extra-galactic particle accelerators.\\ 
The Earth-skimming technique to detect ultra-high energy neutrinos has been proposed before, see for instance \cite{fargion-2,earth-skimming-fargion1, earth-skimming-letessier,earth-skimming-feng, fargion-3}. The method consist on using the interaction between a neutrino and a nucleon via the exchange of a W$^{\pm}$ boson to produce a charged lepton of the same flavour as the incoming neutrino. Since the neutrino-nucleon cross section is very small, a large-mass target is needed. The natural candidate to produce such interactions is the Earth crust, either by focusing the searches on quasi-horizontal neutrinos that travel along a chord inside the Earth or by using mountains as targets. The produced charged lepton travels essentially in the same direction as the neutrino. Thus, in the considered scheme, the charged lepton will travel either upward if it was moving through a chord inside the Earth crust or quasi horizontally if the neutrino passed through a mountain.\\
From the three families of leptons, the electron neutrinos are unfavourable for this type of study because the produced high-energy electrons initiate electromagnetic showers that are easily absorbed by the target mass shortly after production. For $\mu$ neutrinos, the relatively large mean life of the produced charged lepton combined with the ultra-high energy  will produce a detectable signal only as an ultra-energetic $\mu$. The $\tau$ neutrinos are the ones that have attracted more interest from the experimental point of view. The reason is that, because of their very short mean life, even if they are very energetic the produced $\tau$ charged leptons will decay into secondary particles that would make the detection of the signal easier. However, the $\tau$ neutrinos have proven to be one of the most elusive particles of the Standard Model, with less than 15 detections up to now \cite{taudetections}. Ultra-high energy $\tau$ neutrinos (with searches up to 72 $PeV$) have eluded direct detection so far, even after the analysis of three years of IceCube data  \cite{icecube-taus}. Even though $\tau$ neutrinos are disfavoured in production mechanisms at the astrophysical acceleration sites, the neutrino flavour mixing that occurs in cosmological distances is expected to produce approximately equal proportions of all neutrino flavours at the Earth.\\
There have been already experimental attempts to detect ultra-high energy neutrinos using the Earth-skimming technique. For instance, the Pierre Auger observatory in Argentina used their surface detectors to look for the electromagnetic signature of extensive air showers initiated by the decay of $\tau$ charged leptons of $EeV$ energies that develop close to the detector \cite{auger-es-3,auger-es-1,auger-es-2}, without finding candidate signals so far. There are also studies that propose to use the fluorescence detector of the Pierre Auger observatory to detect the decay in the atmosphere of $\tau$ charged leptons produced by ultra-high energy neutrinos \cite{auger-fluor}; however this idea has not been implemented yet. The Ashra-1 collaboration \cite{ashra-es-1} searched for neutrino emission from a GRB in the $PeV$-$EeV$ energy range using the Earth-skimming technique. The Ashra-1 experiment, located on the Mauna Loa volcano and facing the Mauna Kea volcano in the Hawaii island, aimed to detect $\tau$ neutrinos that converted into $\tau$ charged leptons inside Mauna Kea. Their method consisted on measuring the Cherenkov light emitted by the particles of the atmospheric shower initiated by the decay products of the $\tau$ charged lepton. Their analysis of the GRB081203A did not find signals associated to $\tau$ neutrinos in the $PeV$-$EeV$ energy range. There have been also studies about the feasibility to use the MAGIC telescopes for the detection of $\tau$ neutrinos \cite{taus-magic-1,taus-magic-2}, by pointing their telescopes below the horizon towards the sea, or by searching for reflections of Cherenkov light by the nearby ground, the sea or clouds \citep{fargion-5,fargion-4}. However, no experimental results have been published yet.\\
In this paper we propose to adapt the Earth-Skimming technique to use it with the HAWC gamma-ray observatory, employing the Pico de Orizaba volcano as a target for the neutrino-nucleon interactions. However, we propose not to follow the method explored so far of studying the decay products of a $\tau$ charged lepton on the atmosphere. Instead, we propose to reconstruct directly the trajectory of the charged lepton or their decay products as they travel through the HAWC detectors. In this way we do not only restrict our studies to $\tau$ neutrinos, but also include the possibility for the detection of ultra-high energy $\mu$ neutrinos (a first suggestion of this method was mentioned in \cite{stanev}). At ultra-high energies, the produced charged leptons will have an energy approximately equal to that of the original neutrino \cite{earth-skimming-feng,stanev}. These charged leptons, or their decay products, could travel crossing several HAWC detectors depositing large amounts of Cherenkov light, well beyond the average left from both: atmospheric muons ($\approx$ 30 photoelectrons ($PEs$) \cite{AndyCal}) located far from the shower core used in the gamma/hadron discrimination algorithms of HAWC and the high $PE$ noise, also associated to atmospheric muons, considered by the HAWC collaboration to be in the range of 10-200 $PEs$ \cite{Crab}. Here it is important to point out that the dynamical range of the HAWC electronics goes from a fraction of a $PE$ up to thousands $PEs$ \cite{ICRCCal}. 

In order to demonstrate that our proposal to search for tracks produced by $\tau$ charged leptons or their boosted decay products is feasible, we performed some GEANT4 \cite{geant} simulations of the decay of $\tau$ charged leptons with an energy of 1 $PeV$ (that decay approximately 50 $m$ after their creation point), and studied the shower evolution in air. We choose an energy of 1 $PeV$ so the $\tau$ charged leptons decay quickly and are in the energy regime of our studies. After decay, the opening angle of the charged products is smaller than 0.7$^{\circ}$, thus, after the decay products have travelled 2050 $m$ (the approximate distance from the edge of the volcano to the HAWC array at $\approx$ 4090 $m$ a.s.l is of two kilometers), they would hit at most three columns of HAWC tanks. Moreover, over 95\% of the secondary charged particles are contained within an opening angle smaller than 0.2$^{\circ}$, therefore producing large Cherenkov signals only within a single row of HAWC tanks, producing a clearly identifiable track. This simplified exercise was done for 1 $PeV$ charged $\tau$'s, making it easy to extrapolate the results to higher energy charged leptons, since the opening angle of the decay products is inversely proportional to the energy of the primary charged lepton \cite{colimation}.

Based on this information, we believe that the tracking method is possible in the search for charged leptons, produced by neutrino-nucleon interactions, in the $PeV$ energy range. The amount of Cherenkov light that could be detected by the HAWC PMTs by such energetic particles is expected to be in the range of thousands to tens of thousands $PEs$. However, given the dynamic range of the current HAWC electronics one could anticipate that the collected light could be used as a proxy to at least set a lower boundary on the energy of the incoming charged lepton, in a similar manner to what IceCube does for muons that pass through the detector.\\
The paper is organized as follows: in Section \ref{ana-calc} we present the calculation of the flux of charged leptons that could be produced by neutrino-nucleon interactions in the Pico de Orizaba volcano. In Section \ref{ef-area}, we describe a method to calculate the HAWC effective area based on purely geometrical considerations, and evaluate it for different trigger conditions. Then, in Section \ref{results}, we present our results for the possible detection rate and address the issue of the expected background signals. In Section \ref{trigger}, we discuss how the current trigger of HAWC can be useful in selecting data for these studies and discuss further a possible background rejection strategy. Finally, the conclusions are presented in Section \ref{conclusions}.

\section{Calculation of the flux of charged leptons produced by Earth-skimming neutrinos}
\label{ana-calc}

In order to obtain an estimate of the number of ultra-high energy charged leptons that could be produced via the Earth-skimming technique, we follow the formalism developed in \cite{earth-skimming-feng,earth-skimming-feng-pre}. However we use further simplifications due to the detection method that we propose in this paper, and also because of the energy of the neutrinos that we plan to study. The differences between the original formalism of \cite{earth-skimming-feng,earth-skimming-feng-pre} and our implementation are pointed out in the text.\\

The number of charged leptons ($N_{L}$) that could be detected by the observatory is given by

\begin{equation} 
\label{eq1}
N_{L} = \Phi_{L} (A\Omega)_{\mathrm{eff}}TD
\end{equation}

where $\Phi_{L}$ is the charged lepton flux, $(A\Omega)_{\mathrm{eff}}$ is the effective area of the observatory, $T$ is the live time of the experiment and $D$ is the duty cycle for observations, which basically describes which fraction of the time $T$ the experiment is actually able to take data. In this section we describe the calculation of the flux of charged leptons $\Phi_{\mathrm{L}}$ produced by the neutrino-nucleon interactions that occur while the neutrinos traverse the volcano. We start with the differential flux of ultra-high energy neutrinos produced by astrophysical sources, which we consider to be isotropic. This differential flux is given by

\begin{equation} 
\label{eq2}
\frac{d\Phi_{\nu}}{dE_{\nu}dcos\theta_{\nu}d\phi_{\nu}}
\end{equation}

with ($E_{\nu},\theta_{\nu},\phi_{\nu}$) being respectively the energy, polar and azimuthal angle of the neutrinos. Since we are interested on integrating the flux on only a certain region (the one covered by the Pico de Orizaba volcano), and because we are considering an isotropic flux, we can simplify the differential neutrino flux to

\begin{equation} 
\label{eq2b}
\frac{d\Phi_{\nu}}{dE_{\nu}dcos\theta_{\nu}d\phi_{\nu}} = \frac{1}{\Omega} \frac{d\Phi_{\nu}}{\mathrm{dE}_{\nu}}
\end{equation}

where $\Omega$ is the solid angle covered by the volcano that is being used as the target for the neutrino-nucleon interactions. After the proposed isotropic neutrino flux passes through the Pico de Orizaba volcano, the produced differential charged lepton flux is given by

\begin{equation} 
\label{eq3}
\frac{d\Phi_{L}}{dE_{L}dcos\theta_{L}d\phi_{L}}
\end{equation}

with ($E_{L},\theta_{L},\phi_{L}$) being respectively the energy, polar and azimuthal angle of the produced charged leptons. The relation between the differential fluxes of incoming neutrinos and the produced charged leptons is thus given by

\begin{multline}
\label{eq4}
\frac{d\Phi_{L}}{dE_{L}dcos\theta_{L}d\phi_{L}} = \\ \int dE_{\nu} dcos\theta_{\nu}d\phi_{\nu} \frac{1}{\Omega} \frac{d\Phi_{\nu}}{dE_{\nu}} \kappa(E_{\nu},\theta_{\nu}, \phi_{\nu};E_{L},\theta_{L}, \phi_{L})
\end{multline}

where $\kappa$ is a function that physically represents the convolution of the probabilities of the different processes that need to take place in order that an ultra-high energy neutrino converts into a charged lepton inside the target material and is able to escape the mountain. Given that we are interested in studying ultra-high energy neutrinos ($E_{\nu} > 10 \; PeV$), the first simplification comes from the fact that the produced charged lepton will approximately follow the same direction as the original neutrino. The angle between the original neutrino and the produced charged lepton ($\theta_{\nu}-\theta_{L}$) has been estimated to be smaller than 1 $arcmin$ for energies above 1 $PeV$ for $\tau$'s \cite{ashra-es-1}; so, we expect this angle to be negligible. This makes that the angular dependence of $\kappa$ can be approximated by delta functions.

\begin{multline}
\label{eq5}
 \kappa(E_{\nu},\theta_{\nu}, \phi_{\nu};E_{L},\theta_{L}, \phi_{L}) \approx \\ \kappa(E_{\nu};E_{L}) \delta(cos\theta_{\nu}-cos\theta_{L}) \delta(\phi_{\nu}-\phi_{L})
 \end{multline}

The energy dependent part of the $\kappa$ function can be written as the integral along the path of the neutrino, and the corresponding charged lepton, inside the volcano

\begin{equation} 
\label{eq6}
\kappa(E_{\nu};E_{L})  = \int P_{1}P_{2}P_{3}(L)P_{4}
\end{equation}

where $P_{1}$ is the survival probability for a neutrino travelling a certain distance inside the volcano. This probability can be written as

\begin{equation} 
\label{eq7}
P_{1} = \mathrm{exp} \left[ - \int_{0}^{A} \frac{dz'}{L_{CC}^{\nu}(E_{\nu})} \right]
\end{equation}

where $L_{CC}^{\nu}$ is the charged current interaction length. Since the average width of the volcano ($A$) is much smaller than the interaction length, almost all of the neutrinos will traverse the whole mountain. That defines the integration limit of Equation \ref{eq7}.  The interaction length is a function of the density of the medium that the neutrino travels through. In the case of the formalism developed in \cite{earth-skimming-feng}, where the trajectories of the neutrinos were across chords inside Earth, the authors had to consider variations of the Earth density. In this particular case, since we are interested on neutrinos that travel through volcanoes, it is reasonable to consider a constant density $\rho$ along the path of the neutrinos. Thus, the expression for the charged current interaction length can be written as follows

\begin{equation} 
\label{eq8}
L_{CC}^{\nu} (E_{\nu}) = \frac{1}{\sigma_{CC}^{\nu}(E_{\nu})\rho N_{A}} 
\end{equation}

where $N_{A}$ is Avogadro's constant and $\sigma_{CC}(E_{\nu})$ is the charged current cross section. The second probability ($P_{2}$) that enters the calculation of $\kappa$ is that of a neutrino to convert into a charged lepton through a charged current in an infinitesimal distance $dz$. This probability is given by

\begin{equation} 
\label{eq9}
P_{2} = \frac{dz}{L_{CC}^{\nu}(E_{\nu})}
\end{equation}

The third process ($P_{3}$) that is taken into account in the $\kappa$ function is the probability that the produced charged lepton is able to escape the volcano, taking into account the charged lepton energy losses in the medium. This process is described by a system of two coupled differential equations:

\begin{equation} 
\label{eq9b}
\frac{dE_{L}}{dz} = -(\alpha_{L} + \beta_{L}E_{L})\rho
\end{equation}

\begin{equation} 
\label{eq9c}
\frac{dP_{3}(L)}{dz} = -\frac{m_{L}P_{3}(L)}{cT_{L}E_{L}}
\end{equation}

Where Equation \ref{eq9b} describes the energy loss processes, with $\alpha_{L}$ describing the ionization energy loss and $\beta_{L}$ the radiative energy loss. According to the literature, e.g. \cite{earth-skimming-feng}, the effects of $\alpha_{L}$ are negligible at the energy regime of interest of this work; so, we consider in Equation \ref{eq9b} that $\alpha_{L} \rightarrow 0$. In Equation \ref{eq9c}, $m_{L}$ is the mass of the charged lepton, $c$ is the speed of light and $T_{L}$ is the  charged lepton lifetime. A Monte Carlo study from \cite{stanev} shows that for $\mu$'s with energies of 1 $PeV$, $P_{3}(\mu) \approx 0.98$ after 6 $km$ water equivalent ($km.w.e$)  (6 $km.w.e.$ $\approx$ 2.3 $km$ in ``standard rock", the average path length inside the volcano), i.e. in the extreme case where the charged lepton has to travel the average width of the volcano, and this value approaches unity as the energy of the $\mu$ charged lepton increases. Thus, we consider that for the energy regime studied in this work, it is appropriate that for $\mu$'s we can take a value of $P_{3}(\mu) \approx 1.0$ (see Table \ref{tab:tableP3}).

\begin{table}[h!]
  \begin{center}
    \caption{Numerical values of $P_{3}(L)$ for the different energy ranges considered in this work.}
    \label{tab:tableP3}
    \begin{tabular}{ccc}
		\hline	\hline
      Energy bin [$PeV$]  & $P_{3}(\mu)$ & $P_{3}(\tau)$ \\
		\hline
       [$10^{1} , 10^{2}$] & $1$  & $0.5$ \\
       $[10^{2}$, $10^{4}]$  & $1$ & $0.9$ \\
      	\hline \hline
    \end{tabular}
  \end{center}
\end{table}

The case of the survival probability for $\tau$ charged leptons is more complicated to evaluate, since most of the research has been done for energies above or at 100 $PeV$ \cite{tauloss3,tau-energy-loss-2,tauloss4}. We take as a base for our calculations the results presented in \cite{tau-energy-loss-2}. For $E_{\tau} = 10$ $EeV$, $P_{3}(\tau) \approx 0.99$ and for $E_{\tau} = 1$ $EeV$, P$_{3}(\tau) \approx 0.93$ after 2.3 $km$ in ``standard rock", i.e. in the extreme case where the $\tau$'s are produced just after entering the volcano. However, the values of $P_{3}(\tau)$ at lower energies cannot be obtained by a simple extrapolation since the value of $P_{3}(\tau)$ below 100 $PeV$ is dominated by the mean life time of the $\tau$. By considering this fact, we obtain approximate values of $P_{3}(\tau)$ for the $PeV$ energy regime. Table \ref{tab:tableP3} shows the average values of $P_{3}(\tau)$ in the energy range of interest of this work. Finally, the factor $P_{4}$ makes sure that the charged lepton escapes the volume of the volcano with an energy $E_{L}$. Based on Equation \ref{eq9b}, taking $\alpha_{L} \approx 0$, then $P_{4}$ can be written as

\begin{equation} 
\label{eq10}
P_{4} = \delta(E_{L}- E_{\nu} \mathrm{exp}\left[-\beta_{L}\rho z \right] )
\end{equation}

Thus, we can approximate the $\kappa$ function as

\begin{equation} 
\label{eq11}
\kappa(E_{\nu};E_{L})  = \int P_{1}P_{2}P_{3}(L)P_{4} \approx P_{3}(L)\int P_{1}P_{2}P_{4} 
\end{equation}

After evaluating the integral we obtain an expression for the $\kappa$ function. 

\begin{equation} 
\label{eq12}
\kappa(E_{\nu};E_{L})  \approx \frac{P_{3}(L)}{L_{CC}(E_{\nu})}\mathrm{exp}\left[ {\frac{-A}{L_{CC}(E_{\nu})}} \right]\frac{1}{\beta_{L}\rho E_{L}}
\end{equation}

Equation \ref{eq12} is valid in the case where we neglect variations of the density of the volcano. Going back to Equation \ref{eq4}, we can substitute Equation \ref{eq5} and Equation \ref{eq12} on it. By doing this, we obtain an equation for the differential flux with respect to the energy of the leptons

\begin{multline}
\label{eq13}
\frac{d\Phi_{L}}{dE_{L}} = \\ \frac{P_{3}(L)}{\beta_{L}\rho E_{L}} \int dE_{\nu} \frac{d\Phi_{\nu}(E_{\nu})}{dE_{\nu}} \frac{1}{L_{CC}(E_{\nu})}  \mathrm{exp} \left[ -\frac{A}{L_{CC}(E_{\nu})} \right] 
\end{multline}

The next step is to define a function that describes the differential flux of neutrinos as a function of energy. For this, we use the parametrization of the measured astrophysical neutrino flux made by IceCube \cite{IceCubeFlux1,IceCubeFlux2,IceCubeFlux3}, which has been found to be well described by an unbroken power law 

\begin{equation} 
\label{eq14IC}
\frac{d\Phi_{\nu}(E_{\nu})}{dE_{\nu}} = \phi \times \left( \frac{E_{\nu}}{100 \; TeV} \right)^{-\gamma}
\end{equation}

We choose to use the parametrization presented in \cite{IceCubeFlux2} for the measurement of the $\nu_{\mu}+\bar{\nu}_{\mu}$ astrophysical flux. For this particular parametrization we shifted the mean values of the normalization and spectral index in order to obtain the highest flux, within the allowed range given by their statistical uncertainties. These values are: $\phi = 0.3343 \; GeV^{-1} Km^{-2} sr^{-1}$ $yr^{-1}$ and $\gamma = 1.71$. We assume for our calculations an expected equal contribution to the astrophysical flux for $\nu_{\tau}+\bar{\nu}_{\tau}$, due to the neutrino flavour mixing over cosmological distances that would produce approximately equal proportions of all flavours. Then, we extrapolate the measured flux to the energy range 10 $PeV$ to 100 $EeV$. 
Moreover, motivated by the most energetic event found by the diffuse flux muon neutrino search (2009-2015) done by IceCube (a track event that deposited $2.6 \pm 0.3 \; PeV$ in the sensible volume of the detector \cite{UHENeutIC}), we also include in the flux estimations the models proposed in \cite{UHENMod}, which account for sources of multi-$PeV$ neutrinos that are constrained by the most recent ultra-high energy neutrino upper limits set by IceCube \cite{IceCubeFlux4} and Pierre Auger \cite{auger-es-3}. These models, for the sum of all neutrino flavours, have the smoothly-broken power law functional forms

\begin{equation} 
\label{eq15IC}
\frac{d\Phi_{\nu}(E_{\nu})}{dE_{\nu}} = \phi_{i} \times \left[ \left(\frac{E_{\nu}}{E_{i}}\right)^{\alpha\eta} + \left(\frac{E_{\nu}}{E_{i}}\right)^{\beta\eta} \right]^{1/\eta}
\end{equation}

The values of the parameters are: $\alpha=-1$, $\beta=-3$ and $\eta=-1$. $E_{i}$ take the values of: $10^7 \; GeV$ for what we refer in this work as Model A, $10^8 \; GeV$ for Model  B and $10^9 \; GeV$ for Model C. The values of the normalizations $\phi_{i} $ for each of these models are presented in Table \ref{tab:tablem1}

\begin{table}[h!]
  \begin{center}
    \caption{Values of the normalization constants $\phi_{i}$ and pivot energies for the different $PeV$ neutrino models presented in \cite{UHENMod}.}
    \label{tab:tablem1}
    \begin{tabular}{ccc}
		\hline	\hline
      Model  & $E_{i} \; [GeV]$ & $\phi_{i} \; [GeV^{-1} Km^{-2} sr^{-1} yr^{-1}]$ \\
		\hline
      A & $10^{7}$  & $ 6.2545 \times 10^{-5}$ \\
      B  & $10^{8}$ & $1.5780 \times 10^{-6}$ \\
      C & $10^{9}$ & $ 1.5798 \times 10^{-8}$  \\
      	\hline \hline
    \end{tabular}
  \end{center}
\end{table}

According to \cite{UHENMod}, Model A \& B would correspond to the spectra produced by BL Lac AGNs, and combinations of Models A \& C would follow the expected shape of GZK neutrinos produced from EBL and CMB interactions.

In order to calculate the flux of charged leptons produced by neutrino-nucleon interactions we need to define the input parameters that enter the calculation of the number of produced charged leptons (Equation \ref{eq13}). For the value of the parameter that describes the radiative energy losses of the charged leptons as they travel trough the rock ($\beta_{L}$). One can find in the literature $\beta_{\tau}$ parametrizations that may differ by up to a factor of two. For instance, it ranges from 0.26 to 0.59 $\times 10^{-6} \; cm^{2}/g$ at 100 $PeV$ in \cite{tau-energy-loss}. For our calculations we decided to use the $\beta_{\tau}$ values obtained with the ASW structure functions calculated in \cite{tau-energy-loss}; and for $\beta_{\mu}$ the results obtained in \cite{stanev}. For this latter case the results are available up to an energy of 1 $EeV$, but these values are enough for our calculations. For the numerical integrations, we used intermediate values of $\beta_{\tau}$ and $\beta_{\mu}$ in different bins of energy as shown in Table \ref{tab:table1p1}

\begin{table*}
  \begin{center}
    \caption{Numerical values of $\beta_{L}$ calculated with the ASW structure functions from \cite{tau-energy-loss} for $\tau$'s and from the results obtained by \cite{stanev} for $\mu$'s. The table presents the energy at which the parameter was evaluated, and the energy bin in which it is used in the numerical calculations.}
    \label{tab:table1p1}
    \begin{tabular}{cccc}
		\hline	\hline
      $\beta_{L}$ evaluated at [$PeV$] & Energy Bin [$PeV$] & $\beta_{\tau}$ [$cm^{2}/g$] & $\beta_{\mu}$ [$cm^{2}/g$]\\
		\hline
      50 & [$10^{1}$, $10^{2}$] & 2.496 $\times 10^{-7}$ & 4.960 $\times 10^{-6}$ \\
      500 & [$10^{2}$, $10^{3}$] & 2.987 $\times 10^{-7}$ & 5.143 $\times 10^{-6}$ \\
      5 $\times 10^{3}$ & [$10^{3}$, $10^{4}$] & 3.554 $\times 10^{-7}$ & N/A\\
      5 $\times 10^{4}$ & [$10^{4}$, $10^{5}$] & 4.184 $\times 10^{-7}$ & N/A \\	
      	\hline \hline
    \end{tabular}
  \end{center}
\end{table*}

For the density of the volcano, we take that of ``standard rock", $\rho = 2.65\; g/cm^{3}$ \cite{stanev}. For the calculation of the charged current interaction length we use the results from \cite{gandhi}, with the cross section for neutrino-nucleon interactions via charged currents in the energy range $10 \; PeV \le E_{\nu} \le 1 \; ZeV$  given by

\begin{equation} 
\label{eq19}
\sigma_{CC}(\nu N) = 5.53 \times 10^{-36} cm^{2} \left( \frac{E_{\nu}}{1 \; GeV} \right)^{0.363}
\end{equation}

\begin{figure}[h] 
\centering 
\includegraphics[width=0.45\textwidth]{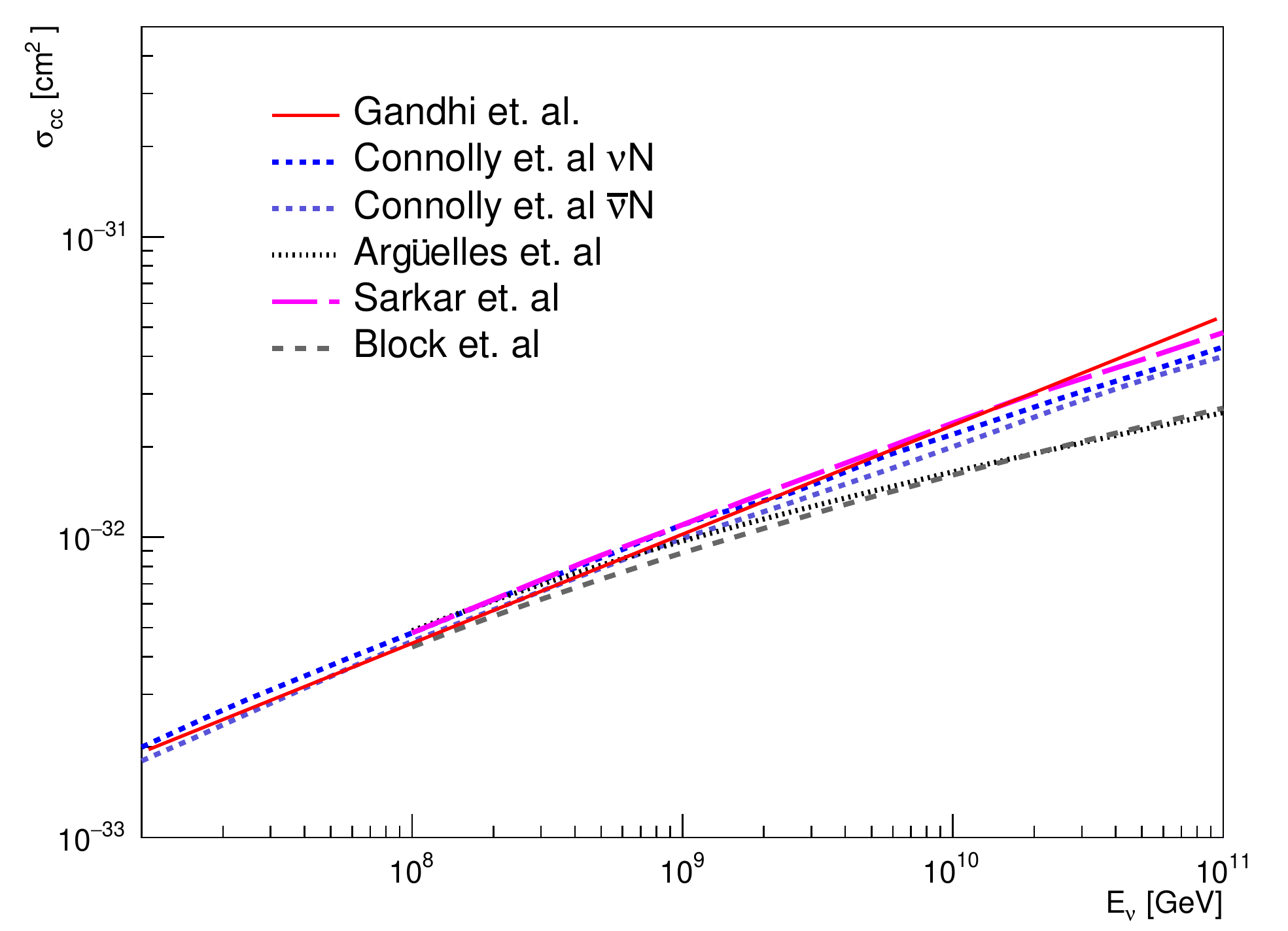} 
\caption{Comparison of the neutrino-nucleon cross section as a function of the neutrino energy, for the model from \cite{gandhi} with those from \cite{conolly} and the results presented in the work of \cite{arguelles}.} 
\label{fig:x-section} 
\end{figure}

This parametrization is a well known result. However, as a cross check, we compared the parametrization given by \cite{gandhi} with more recent calculations \cite{conolly,arguelles}. The result of this is presented in Figure \ref{fig:x-section}. One can notice that the result from \cite{gandhi} follows the general trend predicted by the work of \cite{conolly} and Sarkar et. al. Due to this, and the fact that there is an analytical expression for the cross section of \cite{gandhi}, we decided to use Equation \ref{eq19} in our numerical calculations. The anti neutrino-nucleon cross section is taken to be the same \cite{gandhi}. Using these results we get a value of  $L_{CC}^{\nu} (100 \; PeV) = 1413 \; km$, much smaller than the light-year interaction length for average neutrinos in lead \cite{science-letter-neutrinos}.

For the average width of the Pico de Orizaba, we calculated the typical length of a chord that goes through a cone that follows the geometry of the volcano, as it will be shown in the following section of the paper. At 4100 $m$ a.s.l., the volcano has a width of $\approx$ 6 $km$ and at the summit at $\approx$ 5 500 $m$ a.s.l., a width of $\approx$ 0.35 $km$ (see Figure \ref{fig:pico3} and Figure \ref{fig:pico4}). The average path length for the neutrinos is of 2.33 $km$ inside the volcano. Finally, we assume a duty cycle $D$=95\%, which agrees with the reported value observed during actual HAWC operations \cite{hawc-ani}. The energy range for the charged leptons that exit the volcano is taken in the range ($E_{\mathrm{min}} = \frac{1}{10}  E_{\mathrm{\nu}}$, $E_{\mathrm{max}} =  E_{\mathrm{\nu}}$). In this way we consider an energy range that contains a good fraction of the charged leptons that exit  the volcano. This is because the charged leptons escape with an energy distribution, after loosing some of their energy in the medium. For a study of the propagation of a monoenergetic beam of $\tau$ charged leptons through rock see for instance the work from \cite{tau-energy-loss-2}.

We integrate numerically Equation \ref{eq13} using the differential neutrinos fluxes from Equation \ref{eq14IC} and \ref{eq15IC} to obtain the number of charged leptons ($\Phi_{L}$) that escape the mountain. The results are presented in Tables \ref{tab:tableTauFlux} and \ref{tab:tableMuFlux}.

\begin{table*}
  \begin{center}
    \caption{Tau charged lepton fluxes ($\Phi_{\tau+\bar{\tau}}$) produced by neutrino-nucleon interactions in ``standard rock" with average width $A$. The astrophysical flux is obtained from the extrapolation of the measured neutrino flux by IceCube. See the text for details.}
    \label{tab:tableTauFlux}
    \begin{tabular}{ccccc}
		\hline	\hline
	   	 & \multicolumn{4}{c}{$\Phi_{\tau+\bar{\tau}} \; \; \; [Km^{-2}sr^{-1}yr^{-1}]$}\\
		    \hline
      Neutrino Energy [$E_{\nu}$] & Astrophysical & Model A & Model B & Model C \\
		\hline
      10 $PeV$ - 100 $PeV$ & 10.7471 & 0.4732 & 0.8269 & 0.1008 \\
      100 $PeV$ - 10 $EeV$ & 11.0342 & 0.0233 & 0.4251 & 0.3690 \\
      	\hline \hline
    \end{tabular}
  \end{center}
\end{table*}

\begin{table*}
  \begin{center}
    \caption{Muon charged lepton fluxes ($\Phi_{\mu+\bar{\mu}}$) produced by neutrino-nucleon interactions in ``standard rock" with average width $A$. The astrophysical flux is obtained from the extrapolation of the measured neutrino flux by IceCube. See the text for details.}
    \label{tab:tableMuFlux}
    \begin{tabular}{ccccc}
		\hline	\hline
	   	 & \multicolumn{4}{c}{$\Phi_{\mu+\bar{\mu}} \; \; \; [Km^{-2}sr^{-1}yr^{-1}]$}\\
		    \hline
      Neutrino Energy [$E_{\nu}$] & Astrophysical & Model A & Model B & Model C \\
		\hline
      10 $PeV$ - 100 $PeV$ & 1.0816 & 0.0476 & 0.0832 & 0.0101 \\
      100 $PeV$ - 1 $EeV$ & 0.4685 & 0.0015 & 0.0267 & 0.0185 \\
      	\hline \hline
    \end{tabular}
  \end{center}
\end{table*}

One can notice that the flux of $\mu$ charged leptons is more than an order of magnitude lower than that of $\tau$'s. The reason for this can be seen in Equation \ref{eq13}, where the charged lepton flux depends inversely on the radiative energy losses of the charged leptons, which are more than an order of magnitude higher for the $\mu$'s with respect to $\tau$'s (see Table \ref{tab:table1p1}). This does not contradict the fact that, at the highest energies, for both $\mu$'s and $\tau$'s the survival probability $P_{3}(L)\rightarrow 1$. $P_{3}(L)$ quantifies the probability that the charged leptons are able to escape the volcano, while the $\beta_{L}$ factor in Equation \ref{eq13} appears because of the form of $P_{4}$ and is independent of $P_{3}(L)$. This result is consistent with the arguments developed in \cite{earth-skimming-prl2}, where it is pointed out that $\tau$ charged leptons have a higher probability to escape the Earth crust compared to $\mu$'s.

In the following section we present a simple method that allows to approximate the effective area of the HAWC observatory to ultra-high energy charged leptons.

\section{Effective area determination}
\label{ef-area}

\begin{figure}[h] 
\centering 
\includegraphics[width=0.45\textwidth]{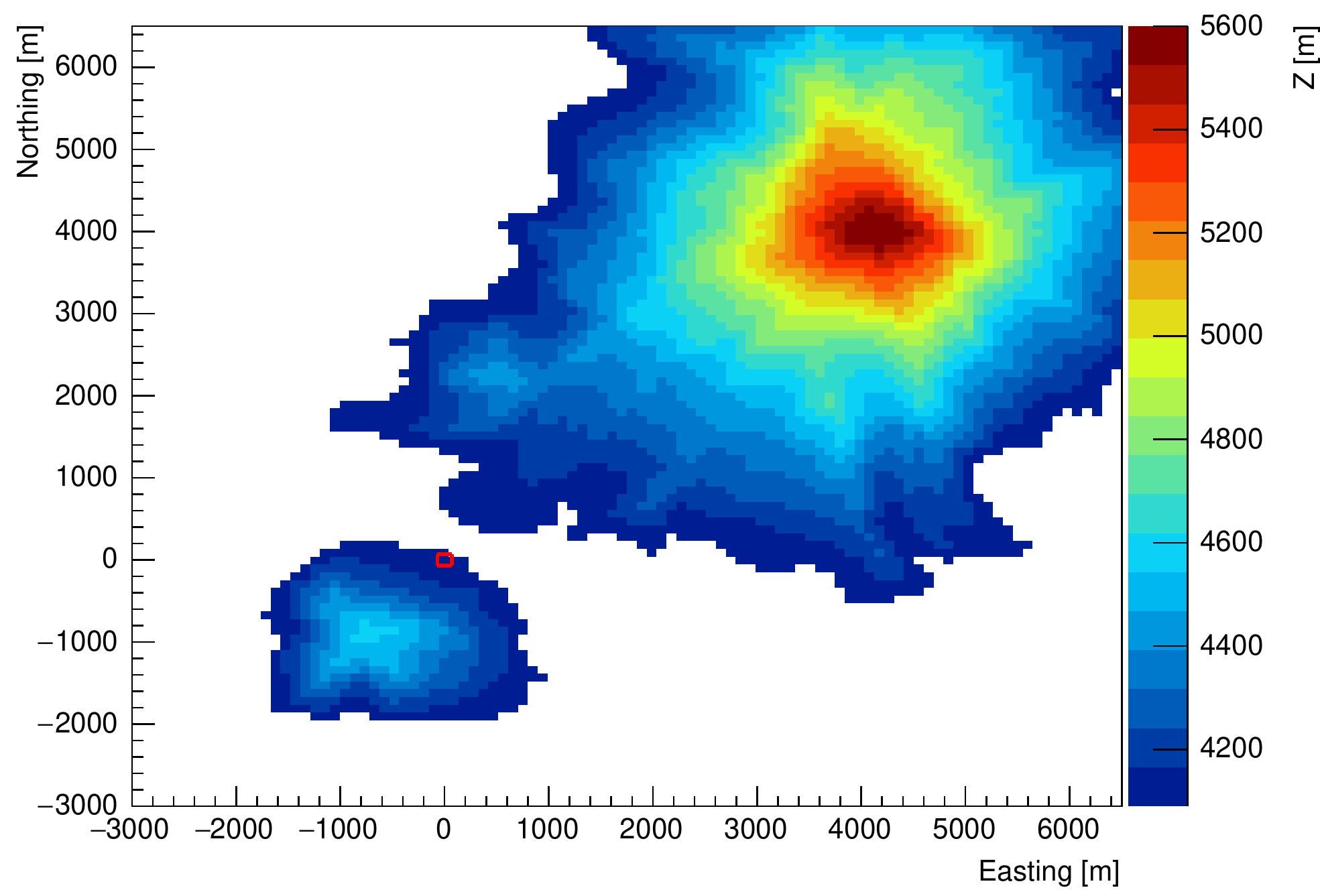} 
\caption{Topography of the HAWC site according to INEGI data. One can see two mountains, the Sierra Negra volcano at bottom left and the much larger Pico de Orizaba volcano on the top right. The small red rectangle, at the origin of the coordinate system, indicates the location and dimensions of the HAWC observatory.} 
\label{fig:pico2} 
\end{figure}

In general, the effective area or acceptance of an observatory is calculated using a detailed Monte Carlo simulation, see for instance in \cite{tau-neutrinos-mc1, eff-area-taus-fullmc} studies of the sensitivity of air Cherenkov and fluorescence observatories to $\tau$ neutrinos using the Earth-skimming technique and in \cite{fargion-5} an early proposal about using fluorescence measurements on observatories as Pierre Auger in order to search for $\tau$ charged lepton showers. The full Monte Carlo method of calculating the effective area represents a complex and time consuming task. In the early results obtained by the Pierre Auger collaboration, their effective area calculation neglected the effect of the topography that surrounds the observatory, and took into account its effect in the systematic error of their observation limits \cite{auger-es-2}. However, their most recent results have incorporated the topography into their effective area calculations in two of their three analysis channels \cite{auger-es-3}. In our case we are specifically interested on the effect of the largest volcano that surrounds the HAWC observatory as a target for the neutrino-nucleon interactions. In order to reproduce the topography that surrounds HAWC, we use data from the \textit{Instituto Nacional de Estad\'istica y Geograf\'ia} (INEGI, M\'exico) \cite{inegi}. Figure \ref{fig:pico2} shows the topography that surrounds the HAWC site (indicated by a small red rectangle). The observatory is located between two volcanoes: in the direction Northing-Easting by the Pico de Orizaba and in the opposite direction by the Sierra Negra volcano. Due to its much larger volume, we focus our attention on the Pico de Orizaba volcano as the target for the neutrino-nucleon interactions.

\begin{figure}[h] 
\centering 
\includegraphics[width=0.45\textwidth]{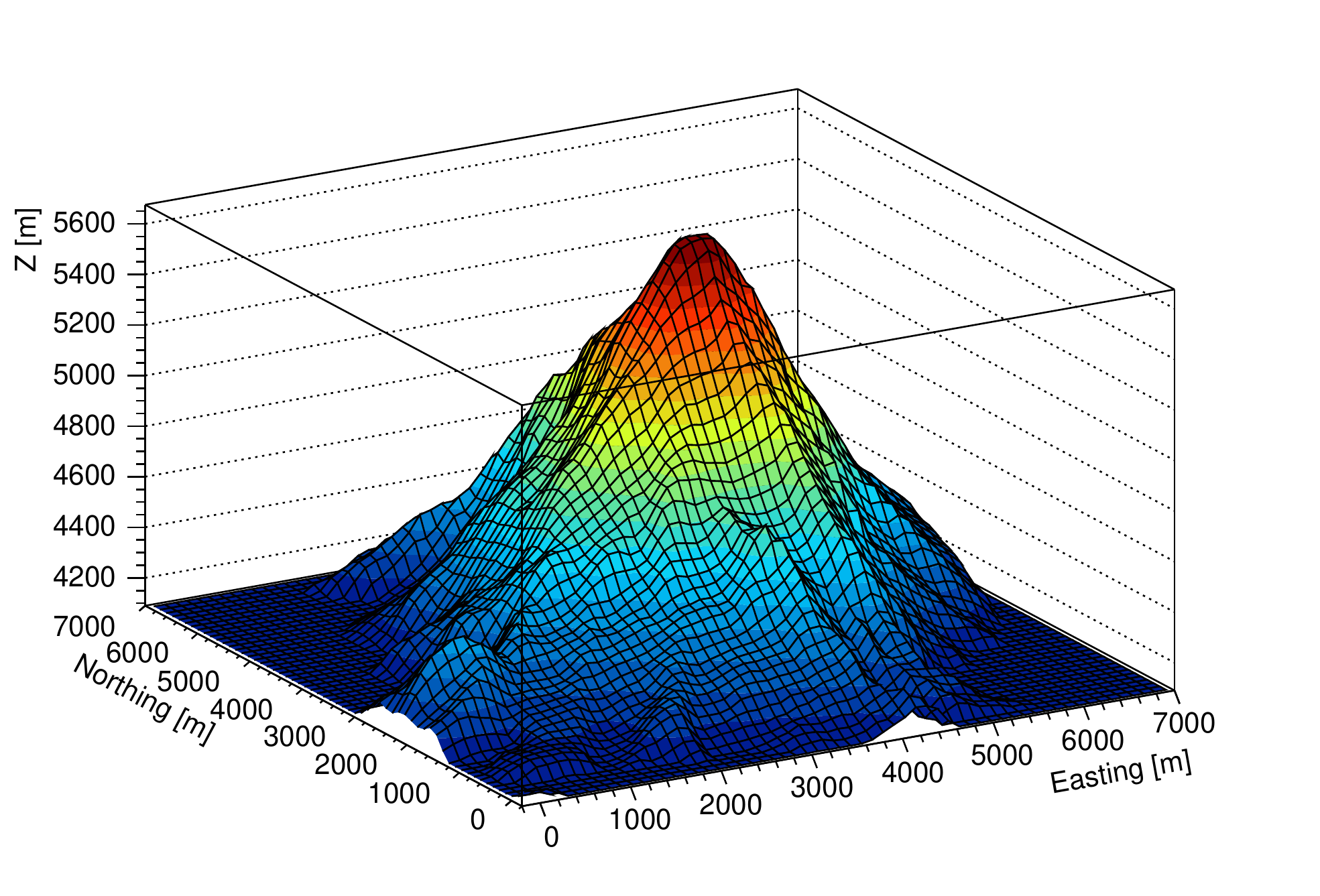} 
\caption{Topography of the Pico de Orizaba volcano, located in the state of Puebla in Mexico. The origin of the coordinate system is centred at the location of the HAWC array.} 
\label{fig:pico1} 
\end{figure}

\begin{figure}[h]
\centering
\begin{minipage}{.45\linewidth}
  \includegraphics[width=\linewidth]{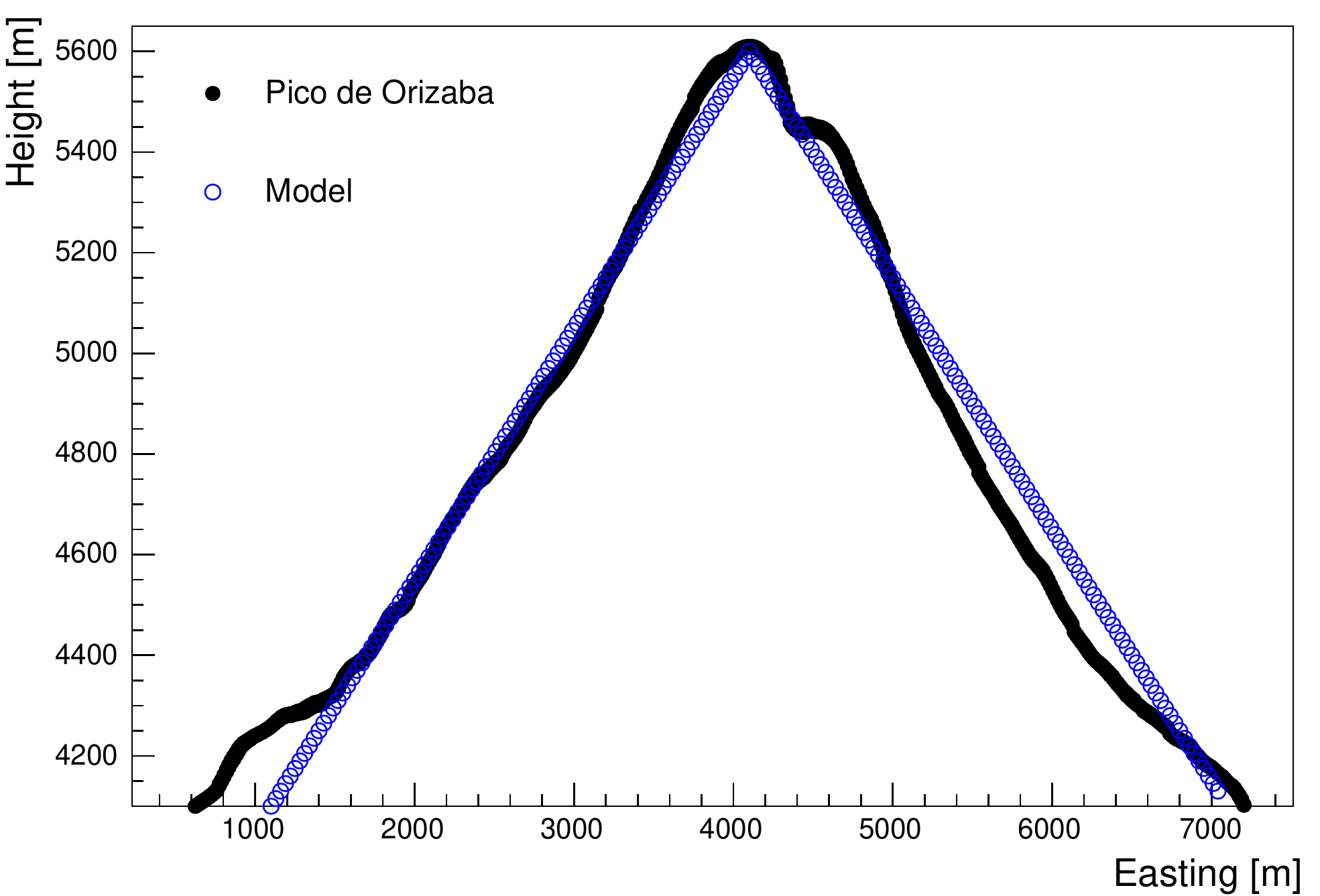}
  \caption{Projection of the Pico de Orizaba profile along the Easting axis. The blue open markers show the profile of the conical approximation to the volcano, while the black closed circles indicate the real profile of the volcano.}
  \label{fig:pico3}
\end{minipage}
\hspace{.05\linewidth}
\begin{minipage}{.45\linewidth}
  \includegraphics[width=\linewidth]{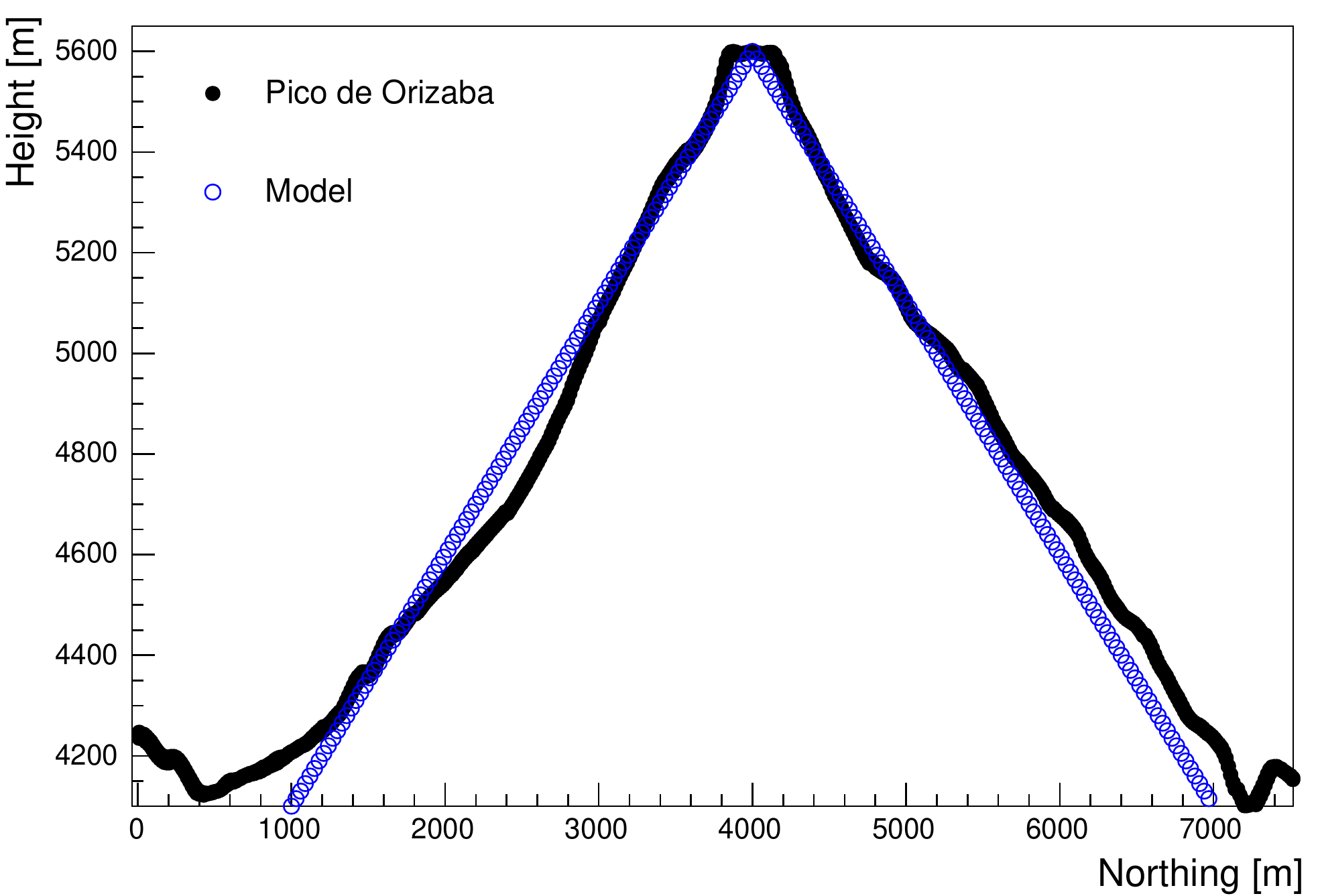}
  \caption{Projection of the Pico de Orizaba profile along the Northing axis. The blue open markers show the profile of the conical approximation to the volcano, while the black closed circles indicate the real profile of the volcano.}
  \label{fig:pico4}
\end{minipage}
\end{figure}

Figure \ref{fig:pico1} shows the topography of the Pico de Orizaba volcano as seen in a coordinate system centred at the location of the HAWC array at $\approx$ 4090 $m$ a.s.l. The approximate symmetry of the Pico de Orizaba volcano makes it easy to motivate a simplification in the calculation of the effective area. We can assume that the geometry of the mountain can be modelled to be conical. Figure \ref{fig:pico3} and Figure \ref{fig:pico4} show the profile of the volcano along the Easting and Northing directions, and how we can approximate this profile using a cone with a height of 1500 m ($z=0$ taken at the altitude above sea level of the HAWC observatory) and a diameter of 6000 m.

\begin{figure}[h]
\centering
\begin{minipage}{.45\linewidth}
  \includegraphics[width=\linewidth]{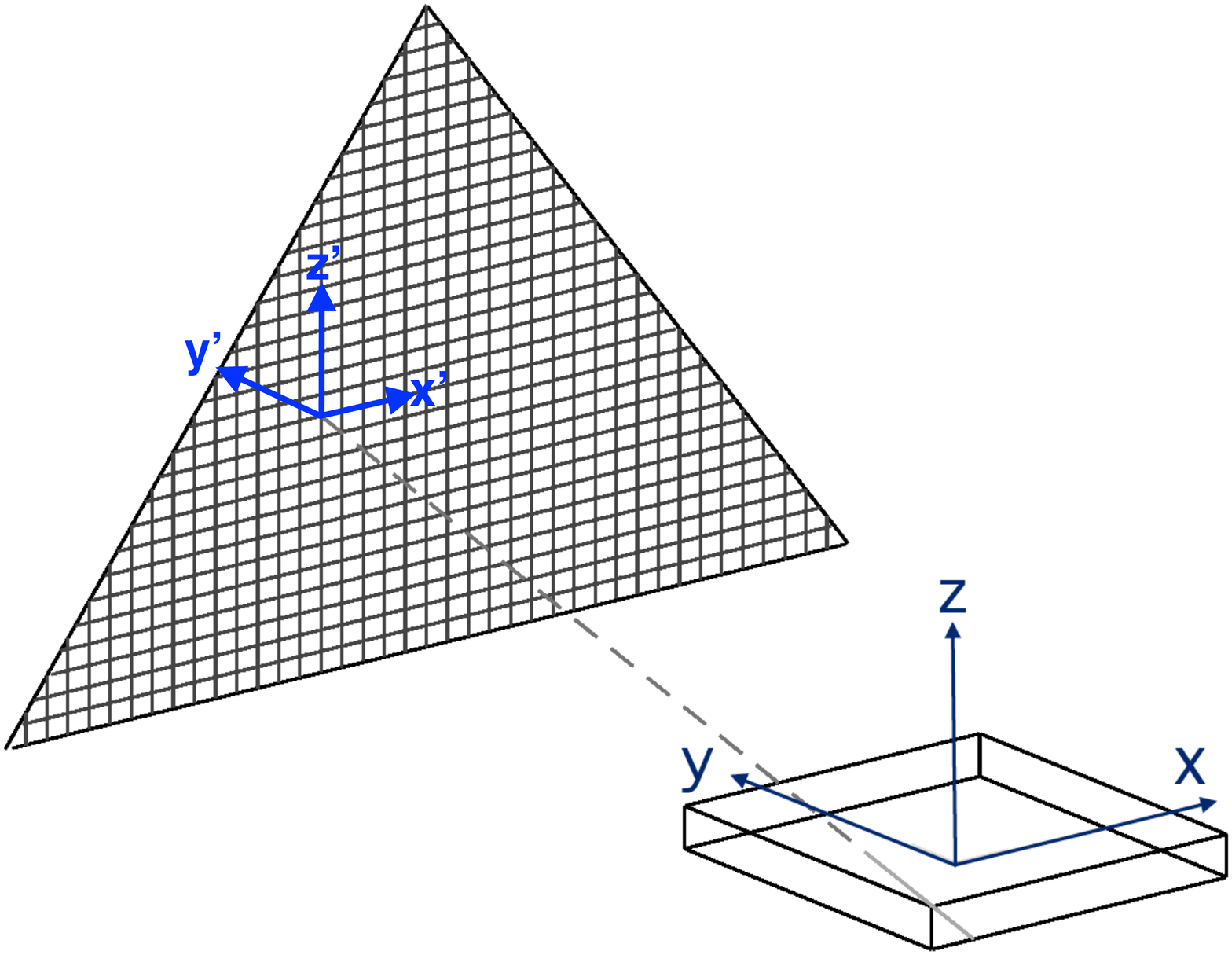}
  \caption{Schematic of the geometry and coordinate system used to simplify the problem. The Pico de Orizaba volcano is represented as a triangular surface located at 5.7 $km$ from the center of the HAWC array. The HAWC detection volume is modelled as a rectangular prism. An example of a vector is shown with a dashed line, that becomes solid as the vector passes through the detection volume. The figure is not drawn to scale.}
  \label{fig:geom}
\end{minipage}
\hspace{.05\linewidth}
\begin{minipage}{.45\linewidth}
  \includegraphics[width=\linewidth]{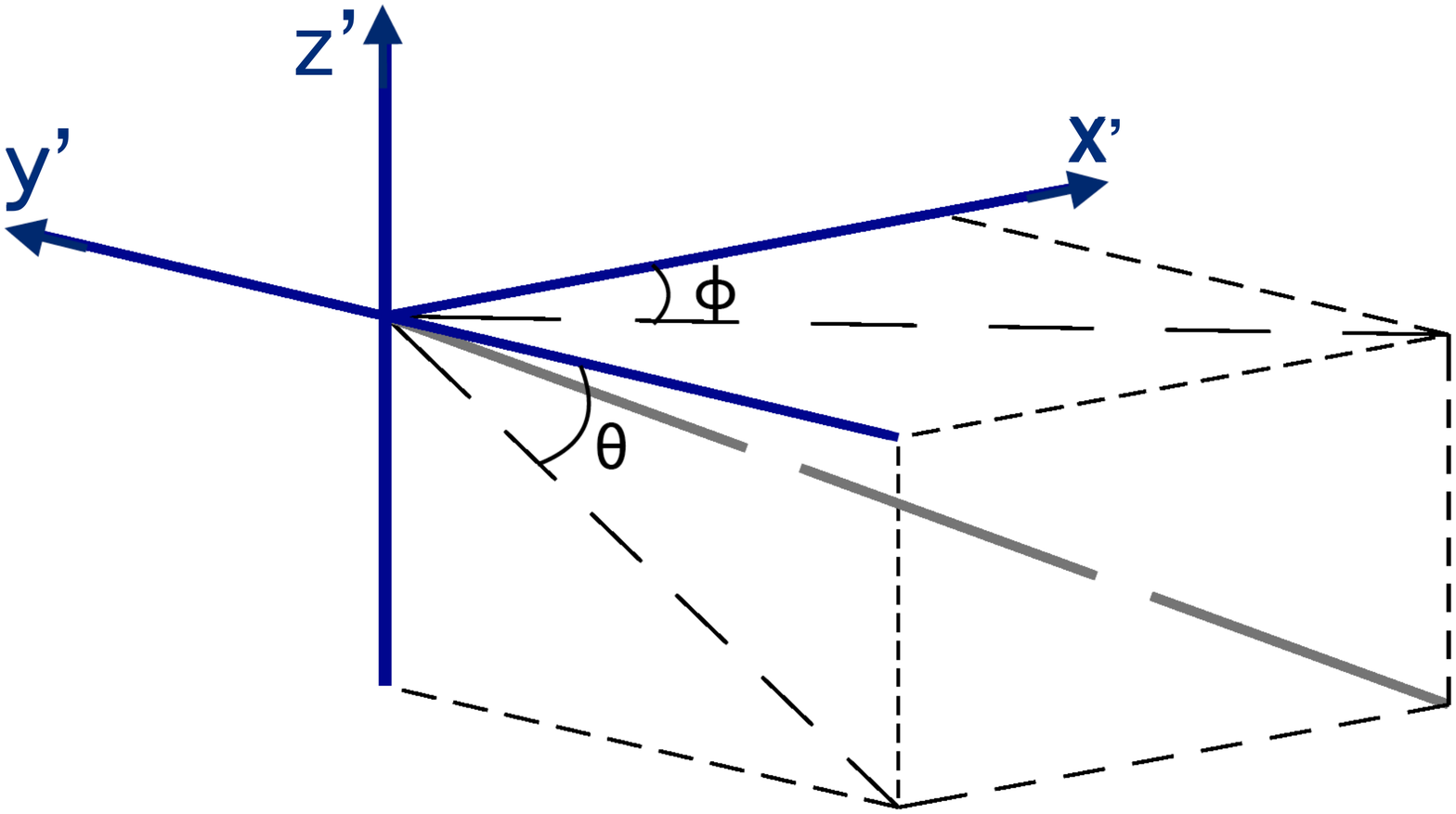}
  \caption{Definition of the angles used in the effective area calculation. The long dashed line represents the orientation of one of the generated vectors. The polar angle $\theta$ is measured between the $-y'$ axis direction and the projection of each vector in the $y'-z'$ plane; it has values equal or greater than $\pi/2$. The azimuthal angle $\phi$ is measured between the $x'$ axis direction and the projection of the vectors in the $x'-y'$ plane; it ranges from 0 to $\pi$.}
  \label{fig:geom2}
\end{minipage}
\end{figure}

Our method to calculate the effective area to detect ultra-high energy charged leptons produced by neutrino-nucleon interaction is as follows:

\begin{enumerate}
\item We consider that the ultra-high energy charged lepton, or their highly boosted decay products, will travel following a straight trajectory along the direction of the initial neutrino.
\item We approximate the region where the charged leptons are produced as a triangular surface, a 2D projection of the volcano, that faces the HAWC observatory. We can use this simplification since here we are only interested on studying the trajectories of the produced charged leptons. The width of the mountain is considered in the charged lepton flux calculation, as presented in Section \ref{ana-calc}. On the triangular surface we draw a grid made of lines parallel to the $x$ \& $z$ axis in the coordinate system shown in Figure \ref{fig:geom}. The lines of the grid have a separation of 0.5 $m$ in each direction. The surface is located at a distance of 5.7 $km$ to the center of the HAWC observatory, approximately the distance between the center of the detector array and that of the volcano.
\item The detection volume of HAWC is modelled as a rectangular prism, with dimensions ($x,y,z$) 140 $m$ $\times$ 140 $m$ $\times$ 4.5 $m$. This approximation is motivated by the actual configuration of the HAWC array. The observatory is made of a compact group of 300 Water Cherenkov Detectors (WCDs). Each WCD is a cylinder having 7.3 $m$ in diameter and 4.5 $m$ in height. The array covers an area of approximately 22 000 $m^{2}$ \cite{hawc-ani}.
\item From each cell of the grid ($\sim14$ million), we generate vectors that point towards our model of HAWC, approximated as a rectangular prism. Figure \ref{fig:geom2} shows a diagram of the definition of the angles used in this work. The azimuth angle $\phi$ covers the range from $0$ to $\pi$ (with steps of 1 degree), pointing towards the -$y'$ direction in the coordinate system shown in Figure \ref{fig:geom}. This is done for the $i$ different orientations of the polar angle $\theta$. The polar angle is defined such that the vector orientation that lies in the $x'-y'$ plane corresponds to 90$^{\circ}$, and increases as the orientation gets closer to the $x'-z'$ plane (see Figure \ref{fig:geom2}).
\item For each orientation $\theta_{i}$, we calculate a differential element of effective area

\begin{equation} 
\label{eq16}
f(\theta_{i}) =A\Delta\phi(\theta_{i}) = \frac{N_{trig}}{N_{gen}} \times A_{i} \times \Delta\phi_{i}
\end{equation}

where, $N_{trig}$ is the total number of vectors whose directions points towards the detection volume of HAWC with a minimum length of the trajectory across the $x-y$ plane shown in Figure \ref{fig:geom} (inside the volume of the rectangular prism). We will refer to this minimum length as the trigger condition, that approximately represents how many WCDs would measure Cherenkov light produced by the incoming lepton or their collimated decay products. $N_{gen}$ is the total number of vectors that point towards the rectangular prism, regardless of the trajectory that they have inside the prism. $A_{i}$ is the section of the area of the triangular surface over which the vectors that point towards the rectangular prism ($N_{gen}$) were generated. Finally, $\Delta\phi_{i}$ is the azimuthal angle covered by the area $A_{i}$, as seen from the center of the rectangular prism.
\end{enumerate}

\begin{figure}[h] 
\centering 
\includegraphics[width=0.45\textwidth]{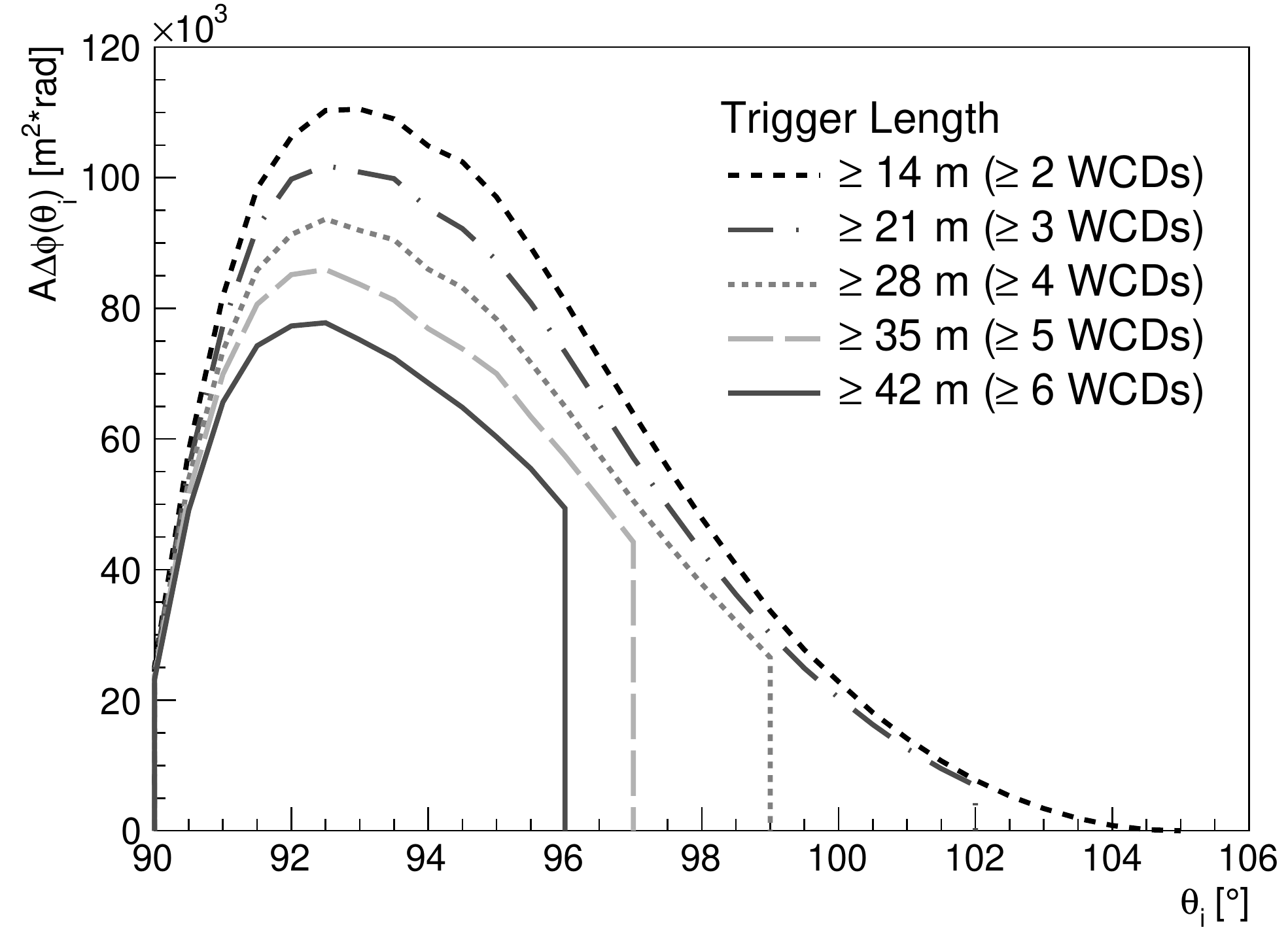} 
\caption{Values of the differential effective area as a function of the polar angle $\theta_{i}$ measured from zenith (the horizontal orientation, parallel to the $y'$ axis, for the vectors is at 90$^{\circ}$). The results are shown for different trigger conditions.} 
\label{fig:eff-area} 
\end{figure}

By following the procedure described above, we can obtain the differential elements of effective area as a function of the polar angle orientation. Figure \ref{fig:eff-area} shows the results for different orientations of the polar angle. 

One can see that the differential effective area increases with increasing values of the polar angle, up to a maximum at $\approx$ 93$^{\circ}$, and then decreases because the area over which the vectors are generated decreases as well, as it corresponds to the upper sections of the triangular grid where the vectors are generated. The different line styles used in the plot show the results obtained for different trigger conditions. For instance, the curve with 14 $m$ as trigger condition indicates the results obtained when the requirement of the path of the charged lepton through the detection volume corresponds to at least a trajectory in the $x-y$ plane inside the HAWC rectangular prism of 14 $m$. This would correspond roughly to have the ultra-high energy charged lepton (or its boosted decay products) to traverse through at least two WCDs. This may not seem enough to even provide a rough idea of the direction of the incoming lepton. However, one should keep in mind that each WCD is equipped with four photomultiplier tubes (PMTs), recording the amplitude and arrival time of the Cherenkov light. By combining the information from the eight PMTs from two WCDs it may be possible to have enough information to reconstruct a trajectory, although details of the angular reconstruction of the tracks is beyond the scope of this work. In Figure \ref{fig:eff-area}, we present the results for trigger conditions that correspond roughly to have the charged lepton (or its collimated decay products) traversing from two  up to six WCDs. As expected, the effective area that corresponds to a given trigger condition decreases as one increases the number of detectors required to have signals. We are interested in the total effective area $(A\Omega)_{\mathrm{eff}}$ that goes into the calculation of the number of possible detections of Equation \ref{eq1}. This total effective area is given by 

\begin{equation} 
\label{eq18}
(A\Omega)_{\mathrm{eff}} = \int_{\theta_{j}}^{\theta_{f}} f(\theta_{i}) d\theta_{i}
\end{equation}

where the integration range in $\theta_{i}$ goes from 90$^{\circ}$ to 105$^{\circ}$ (the angular height of the Pico de Orizaba volcano as seen from the center of the HAWC observatory is of $\approx$ 15$^{\circ}$). Our results of the effective area calculation as a function of the different trigger conditions are shown in Table \ref{tab:table1}.

\begin{table}[h!]
  \begin{center}
    \caption{Results of the effective area calculation for different trigger conditions. The trigger condition is the path length that a charged lepton, or its highly boosted decay products, travel along the $x-y$ plane inside the detection volume of HAWC modelled as a rectangular prism.}
    \label{tab:table1}
    \begin{tabular}{cc}
		\hline	\hline
      Trigger [$m$] & $A\Omega$ [$km^{2} sr$] \\
		\hline
      $\ge$ 14 & 0.013867  \\
      $\ge$ 21 & 0.012544  \\
      $\ge$ 28 & 0.010578  \\
      $\ge$ 35 & 0.008418  \\	
      $\ge$ 42 & 0.006782  \\	
      	\hline \hline
    \end{tabular}
  \end{center}
\end{table}
 
With the results from Sections \ref{ana-calc} and \ref{ef-area}  we have: the flux of charged leptons given a certain isotropic flux of neutrinos and the effective area of the HAWC observatory obtained with a simple trigger condition. Using this, in the next section we present our results for the flux of ultra-high energy charged leptons that could be detected by the HAWC observatory.

\section{Results, possible background signals and discussion}
\label{results}

Table \ref{tab:table1mu} shows the number of detectable charged $\tau$'s produced by neutrino-nucleon interactions in the Pico de Orizaba volcano, obtained using the effective area results presented in Table \ref{tab:table1} and the ultra-high energy $\tau$ lepton fluxes from Table \ref{tab:tableTauFlux}. We restrict the results to those from $\tau$ charged leptons since, as it was shown in Section \ref{ana-calc}, the flux from $\mu$ charged leptons is approximately an order of magnitude lower, making their detection not feasible with the method proposed in this work. The detection estimates are shown as a function of the different trigger conditions and of the different ultra-high energy neutrino fluxes discussed in Section \ref{ana-calc}. The eight column presents an estimate of the background signals that are expected using this method. We estimated the background considering that the main contribution is that of ultra-high energy atmospheric muons, coming from the direction of the Pico de Orizaba. In order to calculate this, we used the characterization of the atmospheric muon flux above 15 $TeV$ measured by IceCube \cite{atmos-mu-ic}, that can be modelled as an unbroken power law:

\begin{equation} 
\label{eq-bckg}
\frac{d\Phi_{\mu}}{dE_{\mu}} = \phi_{\mu} \times \left( \frac{E_{\mu}}{10 \; TeV} \right)^{-\delta}
\end{equation}

In this case, we shifted the mean values (within the statistical and systematic uncertainties of the IceCube measurement) of the normalization and spectral index in order to obtain the worst case scenario of the background, and then extrapolated the muon flux to the energy range of our interest. The values of the parameters that maximize the background muon flux are: $\phi_{\mu} = 4.6673 \times 10^{7} \; TeV^{-1} Km^{-2} sr^{-1}yr^{-1}$ and a spectral index of 3.73. Thus, the background flux can be calculated using Equation \ref{eq1}, where in this case the charged lepton flux corresponds to the integral of Equation \ref{eq-bckg} within the energy bins used in this work. Of course this is an overestimation, since these atmospheric muons would have to be able to survive their travel through the volcano, however we consider this estimation of the background appropriate because the study of the propagation of atmospheric muons through the volcano is beyond the scope of this work. Moreover, it has already been noticed that the atmospheric muon background is relevant up to energies of $\approx$ 100 $TeV$ \cite{atmos-mu-2}. The last two columns of Table \ref{tab:table1mu} show the estimated time to get a detection for each trigger condition, first for the case of only having as a source the neutrino flux given by the extrapolation of the measurement done by IceCube \cite{IceCubeFlux1}, and also for the case of including GZK neutrinos from \cite{UHENMod}. \\
From Table \ref{tab:table1mu}, we find that it would take around eight years of data taking in order to be able to find a track-like signal traversing three WCDs based on the extrapolation of the measured astrophysical flux, for $\tau$ neutrinos in the energy range from 10 $PeV$ to 100 $PeV$. In order to find a track-like signal that propagates through four WCDs it could take more than nine years of data taking, but such a long track may allow to reconstruct with better accuracy the direction of the original neutrino. We have verified the predicted detection rate by doing a back of the envelope comparison with the $PeV$ neutrino measurements reported by IceCube \cite{ice-cube1}, and found that, taking into account the different experimental conditions, our results fall in the expected order of magnitude.\\
The HAWC observatory is currently planned to operate for at least five years \cite{hawc-5}, but given its importance as a trigger for future instruments such as CTA, it is not unlikely that it may operate for at least 10 years. Table \ref{tab:table1mu} shows also for completeness the results expected for neutrinos in the energy range  from 100 $PeV$ to 100 $EeV$. Charged $\tau$ leptons that do not decay before arriving or inside the HAWC array (energies of $\approx$ 100 $PeV$ or larger) will leave almost the same signal because the values of $\beta_{L}$ (see Table \ref{tab:table1p1}) change very slowly with energy \citep{fargion-3}. This condition will make very difficult to estimate the energy of the incoming $\tau$. The results also show that the background from ultra-high energy muons at the energy regime of interest of this work is very low and would allow a clean identification of a signal coming from the direction of the Pico de Orizaba volcano.

\begin{sidewaystable*} 
  \begin{center}
    \caption{Number of detectable $\tau$ charged leptons per year ($N$) in HAWC for two different energy bins: (10 $PeV$-100 $PeV$) and (100 $PeV$-100 $EeV$) for the extrapolation of the measured IceCube (IC) flux \cite{IceCubeFlux1}, and for the ultra-high energy predictions ($\Phi_{A},\Phi_{B}$ and $\Phi_{C}$ for the models A, B and C discussed in Section \ref{ana-calc}) from \cite{UHENMod}. The results are presented for different trigger conditions that corresponds to how many HAWC WCDs participate in a track-like event. The addition of the flux $\Phi_{A}+\Phi_{C}$ to the extrapolation of the IC flux is motivated by their approximation to GZK neutrinos produced by interactions with the EBL and CMB \cite{UHENMod}. The eighth column (Bckg) shows the expected number of background signals per year, see the text for details. The last two columns show the time needed (in years) for one charged $\tau$ lepton detection ($\Delta$T), using the extrapolation of the IceCube measured flux and also considering the additional GZK flux at ultra-high energy proposed in \cite{UHENMod}.}
    \label{tab:table1mu}
    \begin{tabular}{cccccccccc}
		\hline	\hline
      Neutrino energy [$E_{\nu}$] & Trigger [$m$] & $N$:IC & $N$:$\Phi_{A}$ & $N$:$\Phi_{B}$ & $N$:$\Phi_{C}$ & $N$:IC $+ \Phi_{A} + \Phi_{C}$ & Bckg & $\Delta$T IC [$yr$] & $\Delta$T IC $+ \Phi_{A} + \Phi_{C}$ [$yr$]\\
		\hline
   \multirow{5}{*}{10 $PeV$ - 100 $PeV$} & $\ge$ 14 & 0.1416 & 0.0063 & 0.0109 & 0.0014 & 0.1492 &  0.0145 & 7.1 & 6.7 \\
    & $\ge$ 21 & 0.1281 & 0.0057 & 0.0099 & 0.0012 & 0.1349 & 0.0131 & 7.8 & 7.4   \\
    & $\ge$ 28 & 0.1080 & 0.0048 & 0.0083 & 0.0010 &  0.1138 & 0.0111 & 9.3 & 8.8  \\
    & $\ge$ 35 & 0.0860 & 0.0038 & 0.0066 & 0.0008 & 0.0906 & 0.0088 & 11.6 & 11.0  \\	
    &  $\ge$ 42 & 0.0693 & 0.0031 & 0.0054 & 0.0007 & 0.0730 & 0.0071 & 14.4 & 13.7 \\	
      	\hline \hline
   \multirow{5}{*}{100 $PeV$ - 100 $EeV$} & $\ge$ 14 & 0.1454 & 0.0003 & 0.0056 & 0.0049 & 0.1505 & $ < 10^{-5}$ & 6.9 & 6.6 \\
    & $\ge$ 21 & 0.1315 & 0.0003 & 0.0050 & 0.0044 & 0.1321 & $ < 10^{-5}$ & 7.6 & 7.6 \\
    & $\ge$ 28 & 0.1109 & 0.0003 & 0.0042 & 0.0037 & 0.1148 & $ < 10^{-5}$ & 9.0 & 8.7 \\
    & $\ge$ 35 & 0.0882 & 0.0002 & 0.0034 & 0.0030 & 0.0914 & $ < 10^{-5}$ & 11.3 & 10.9 \\	
    &  $\ge$ 42 & 0.0711 & 0.0002 & 0.0027 & 0.0023 & 0.0736 & $ < 10^{-5}$ & 14.1 & 13.6 \\	
      	\hline \hline      	
    \end{tabular}
  \end{center}
\end{sidewaystable*} 

\section{Triggering of signals and background rejection}
\label{trigger}

The current software trigger used by the HAWC observatory consists on requiring a certain number of PMTs to be above a charge threshold (it actually consist on two thresholds used to estimate the charge with the Time-over-Threshold method; the thresholds are of 0.25 and 4 $PEs$) within a time sliding window of the order of $\approx$ 150 $ns$ \cite{AndyCal}. For the HAWC results presented in \cite{hawc-ani}, the number of PMTs required to trigger an event was of 15. Thus, the multiplicity trigger used by the HAWC Collaboration is already useful for the collection of signals that are needed for the neutrino searches proposed in this work. To show this, we can estimate the average number of PMTs that have signals during any 150 $ns$ time window. From \cite{hawc-grbs} we know that the sum of the signals from 112 HAWC's PMTs in a 60 $s$ time window was of $\approx$ 2.285$\times 10^{8}$. Then, we can use this number to estimate the number of PMTs fired in any 150 $ns$ triggering window, which is of $\approx$ 6. Thus, only 9 additional PMTs would be required to have signals within the specified time window in order to produce a trigger useful for the neutrino search proposed in this work. This condition can be easily fulfilled if an ultra-high energy lepton or its decay products pass through 3 WCDs, firing at least nine of its 12 PMTs (with a propagation time of the corresponding signals along 3 WCDs of $\approx$ 70 $ns$, well within the current trigger window). Of course, a specialized trigger can be developed in order to obtain a better trigger efficiency. This could be done, for instance, by selecting events with topologies consistent with tracks propagating through the WCDs. This would also have the consequence of reducing the bandwidth needed for this additional trigger.

A second point of interest to discuss is the ability of the instrument to separate the signals produced by ultra-high energy charged leptons from those from the background of lower energy atmospheric muons. The amount of emitted Cherenkov photons per unit length inside a HAWC WCD is given by the Frank-Tamm equation; and the total amount of light will be proportional to the the sum of the path lengths of the primary and all the produced secondary charged particles (from bremsstrahlung and pair production processes) that travel inside a WCD and have a velocity larger than the threshold for Cherenkov light production. According to the study presented in \cite{CLight} this sum of path lengths increases linearly with the energy of the primary particle. Thus, the way to discriminate background signals could be based on the total light yield detected in the WCDs that belong to a given track (e.g. a 1 $TeV$ primary electron would produce an order of magnitude more photons than those from a 100 $GeV$ primary electron). Based on the results presented in \cite{pdg}, the mean momentum of nearly horizontal tracks ($\theta$ =  75 $^{\circ}$) is of $\approx$ 200 $GeV$/c, and in this work we aim to detect the tracks produced by leptons of $PeV$ energies, so we could expect a factor larger than $10^3$ in the deposited light in the WCDs for the ultra-high energy neutrino initiated signals relative to the average muon background tracks. Moreover, the results presented in \cite{muonshz} show that the integral intensity of muons of $\approx$ 1 $TeV$/c is larger than that of $\approx$ 200 $GeV$/c muons by a factor of $\approx$ 52, making reasonable to easily handle the data rate if a specialized trigger is set to keep the data of tracks related to large deposits of Cherenkov light.  \\
An additional experimental proof that this analysis could be performed, is the ability to observe the cosmic ray shadow produced by the volcano, as for instance pointed out in \cite{fargion-5}. However, this requires the use of data and algorithms property of the HAWC Collaboration that are no publicly available. Nonetheless we are confident, because of the interest on this result, that the Pico de Orizaba cosmic ray shadow will be publicly available soon.\\ 
Due to this facts, we believe that it is feasible to separate the huge atmospheric background from the signals we are looking for. We are aware that the precision that could be achieve in determining the energy of the primary tau lepton will be low, nonetheless the primary interest of this work is the detection technique and we leave the development of a possible energy estimator for future work.

\section{Conclusions}
\label{conclusions}

We presented an estimate of the detection capabilities of the HAWC observatory to study ultra-high energy neutrinos interacting in the Pico de Orizaba volcano. We based our study in the analytic method developed by \cite{earth-skimming-feng,earth-skimming-feng-pre} and modified it to a simpler case where the neutrino conversion takes place inside a medium of constant density. The effective area of the HAWC observatory for the detection of ultra-high energy charged leptons was calculated geometrically for different trigger conditions. We used the astrophysical neutrino flux measured by IceCube \cite{IceCubeFlux1} and extrapolated it to energy range from 10 $PeV$ up to 100 $EeV$, and also considered models for multi-$PeV$ neutrinos \cite{UHENMod} that are constrained by the most recent data from both IceCube \cite{IceCubeFlux4} and the Pierre Auger observatory \cite{auger-es-3}. With this, we found that in order to find a signal consistent with the propagation of an ultra-high energy charged lepton coming from the direction of the Pico de Orizaba volcano, will require approximately nine years of data taking if the signal is required to propagate in four WCDs of HAWC. We estimated the expected background for this analysis using the measured atmospheric muon flux above 15 $TeV$ \cite{atmos-mu-ic}, and found that it should be feasible to perform this study with a reasonable signal to background ratio during the lifetime of the HAWC observatory. We also showed that the current software trigger used by HAWC should be sufficient to acquire data for this analysis, and that given the dynamic range of the HAWC electronics it is feasible to be able to discriminate the background signals produced by lower energy muons using the light yield detected by the HAWC PMTs.\\
As an anonymous referee pointed out to us, the detection rate should be taken with caution since there are several quantities that are uncertain to at least some degree, such as the flux of neutrinos at the ultra-high energy regime, the neutrino-nucleon cross section, and the parameters that describes the radiative energy loss processes for the leptons traversing the Earth crust. By selecting a different set of the input parameters for the calculations one could find a detection rate that could decrease by a factor of three or even more. However we also did not consider scenarios where the neutrino-nucleon cross section at ultra-high energies could have enhanced values relative to the standard model predictions, due to the presence of new physics, as described for instance by \cite{NXs-Sarcevic}. Further work should be done using a complete Monte Carlo simulation that incorporates the detailed topography of the volcanoes that surround the HAWC observatory and the complete detector simulation to study the detector response. However this first step shows encouraging results to pursue more detailed studies. The detection rate is certainly low, not comparable to the one that dedicated neutrino experiment can achieve, but as pointed out in \cite{UHENMod}, a single detection of a neutrino with an energy higher than 10 $PeV$, would give evidence of a flux beyond what is firmly established. Our results indicate that, although extremely challenging, it is worth trying to detect ultra-high energy neutrinos, interacting in the Pico de Orizaba with the HAWC observatory.


\section*{Competing Interests}
The authors declare that there is no conflict of interest regarding the publication of this article.

\section*{Acknowledgments}
The authors thank an anonymous referee for very useful discussions that allowed to greatly improve the quality of this work. This work was supported by the programme UNAM-DGAPA-PAPIIT  IA102715, Consejo Nacional de Ciencia y Tecnolog\'ia (CONACyT), Mexico grant 254964 and Coordinaci\'on de la Investigaci\'on Cient\'ifica UNAM.

%
%

\nocite{*}
\bibliographystyle{unsrtnat} 
\bibliography{biblio-hlv2.bib}

\begin{thebibliography}{53}
\providecommand{\natexlab}[1]{#1}
\providecommand{\url}[1]{\texttt{#1}}
\expandafter\ifx\csname urlstyle\endcsname\relax
  \providecommand{\doi}[1]{doi: #1}\else
  \providecommand{\doi}{doi: \begingroup \urlstyle{rm}\Url}\fi

\bibitem[{IceCube Collaboration}(2013)]{ice-cube1}
{IceCube Collaboration}.
\newblock {Evidence for High-Energy Extraterrestrial Neutrinos at the IceCube
  Detector}.
\newblock \emph{Science}, 342:\penalty0 1242856, November 2013.

\bibitem[{Abeysekara} et~al.(2013){Abeysekara}, {Alfaro}, {Alvarez},
  {{\'A}lvarez}, {Arceo}, {Arteaga-Vel{\'a}zquez}, {Ayala Solares}, {Barber},
  {Baughman}, {Bautista-Elivar}, {Belmont}, {BenZvi}, {Berley}, {Bonilla
  Rosales}, {Braun}, {Caballero-Lopez}, {Carrami{\~n}ana}, {Castillo}, {Cotti},
  {Cotzomi}, {de la Fuente}, {De Le{\'o}n}, {DeYoung}, {Diaz Hernandez},
  {Diaz-Velez}, {Dingus}, {DuVernois}, {Ellsworth}, {Fernandez}, {Fiorino},
  {Fraija}, {Galindo}, {Garcia-Luna}, {Garcia-Torales}, {Garfias},
  {Gonz{\'a}lez}, {Gonz{\'a}lez}, {Goodman}, {Grabski}, {Gussert},
  {Hampel-Arias}, {Hui}, {H{\"u}ntemeyer}, {Imran}, {Iriarte}, {Karn}, {Kieda},
  {Kunde}, {Lara}, {Lauer}, {Lee}, {Lennarz}, {Le{\'o}n Vargas}, {Linares},
  {Linnemann}, {Longo}, {Luna-Garc{\'{\i}}a}, {Marinelli}, {Martinez},
  {Mart{\'{\i}}nez-Castro}, {Matthews}, {Miranda-Romagnoli}, {Moreno},
  {Mostaf{\'a}}, {Nava}, {Nellen}, {Newbold}, {Noriega-Papaqui},
  {Oceguera-Becerra}, {Patricelli}, {Pelayo}, {P{\'e}rez-P{\'e}rez}, {Pretz},
  {Rivi{\`e}re}, {Ryan}, {Rosa-Gonz{\'a}lez}, {Salazar}, {Salesa}, {Sandoval},
  {Santos}, {Schneider}, {Silich}, {Sinnis}, {Smith}, {Sparks}, {Springer},
  {Taboada}, {Toale}, {Tollefson}, {Torres}, {Ukwatta}, {Villase{\~n}or},
  {Weisgarber}, {Westerhoff}, {Wisher}, {Wood}, {Yodh}, {Younk}, {Zaborov},
  {Zepeda}, and {Zhou}]{hawc-gamma-rays}
A.~U. {Abeysekara}, R.~{Alfaro}, C.~{Alvarez}, J.~D. {{\'A}lvarez}, R.~{Arceo},
  J.~C. {Arteaga-Vel{\'a}zquez}, H.~A. {Ayala Solares}, A.~S. {Barber}, B.~M.
  {Baughman}, N.~{Bautista-Elivar}, E.~{Belmont}, S.~Y. {BenZvi}, D.~{Berley},
  M.~{Bonilla Rosales}, J.~{Braun}, R.~A. {Caballero-Lopez},
  A.~{Carrami{\~n}ana}, M.~{Castillo}, U.~{Cotti}, J.~{Cotzomi}, E.~{de la
  Fuente}, C.~{De Le{\'o}n}, T.~{DeYoung}, R.~{Diaz Hernandez}, J.~C.
  {Diaz-Velez}, B.~L. {Dingus}, M.~A. {DuVernois}, R.~W. {Ellsworth},
  A.~{Fernandez}, D.~W. {Fiorino}, N.~{Fraija}, A.~{Galindo}, J.~L.
  {Garcia-Luna}, G.~{Garcia-Torales}, F.~{Garfias}, L.~X. {Gonz{\'a}lez}, M.~M.
  {Gonz{\'a}lez}, J.~A. {Goodman}, V.~{Grabski}, M.~{Gussert},
  Z.~{Hampel-Arias}, C.~M. {Hui}, P.~{H{\"u}ntemeyer}, A.~{Imran},
  A.~{Iriarte}, P.~{Karn}, D.~{Kieda}, G.~J. {Kunde}, A.~{Lara}, R.~J. {Lauer},
  W.~H. {Lee}, D.~{Lennarz}, H.~{Le{\'o}n Vargas}, E.~C. {Linares}, J.~T.
  {Linnemann}, M.~{Longo}, R.~{Luna-Garc{\'{\i}}a}, A.~{Marinelli},
  O.~{Martinez}, J.~{Mart{\'{\i}}nez-Castro}, J.~A.~J. {Matthews},
  P.~{Miranda-Romagnoli}, E.~{Moreno}, M.~{Mostaf{\'a}}, J.~{Nava},
  L.~{Nellen}, M.~{Newbold}, R.~{Noriega-Papaqui}, T.~{Oceguera-Becerra},
  B.~{Patricelli}, R.~{Pelayo}, E.~G. {P{\'e}rez-P{\'e}rez}, J.~{Pretz},
  C.~{Rivi{\`e}re}, J.~{Ryan}, D.~{Rosa-Gonz{\'a}lez}, H.~{Salazar},
  F.~{Salesa}, A.~{Sandoval}, E.~{Santos}, M.~{Schneider}, S.~{Silich},
  G.~{Sinnis}, A.~J. {Smith}, K.~{Sparks}, R.~W. {Springer}, I.~{Taboada},
  P.~A. {Toale}, K.~{Tollefson}, I.~{Torres}, T.~N. {Ukwatta},
  L.~{Villase{\~n}or}, T.~{Weisgarber}, S.~{Westerhoff}, I.~G. {Wisher},
  J.~{Wood}, G.~B. {Yodh}, P.~W. {Younk}, D.~{Zaborov}, A.~{Zepeda}, and
  H.~{Zhou}.
\newblock {Sensitivity of the high altitude water Cherenkov detector to sources
  of multi-TeV gamma rays}.
\newblock \emph{Astroparticle Physics}, 50:\penalty0 26--32, December 2013.

\bibitem[Gould and Schr\'eder(1966)]{opacity}
Robert~J. Gould and Gerald Schr\'eder.
\newblock Opacity of the universe to high-energy photons.
\newblock \emph{Phys. Rev. Lett.}, 16:\penalty0 252--254, Feb 1966.

\bibitem[{Fargion} et~al.(1999){Fargion}, A., and R.]{fargion-2}
D.~{Fargion}, {Aiello} A., and {Conversano} R.
\newblock {Horizontal Tau air showers from mountains in deep vally :Traces of
  Ultrahigh neutrino tau}.
\newblock \emph{International Cosmic Ray Conference}, 2:\penalty0 396, August
  1999.

\bibitem[Fargion(2002)]{earth-skimming-fargion1}
D.~Fargion.
\newblock Discovering ultra-high-energy neutrinos through horizontal and upward
  τ air showers: Evidence in terrestrial gamma flashes?
\newblock \emph{The Astrophysical Journal}, 570\penalty0 (2):\penalty0 909,
  2002.

\bibitem[Letessier-Selvon(2001)]{earth-skimming-letessier}
A.~Letessier-Selvon.
\newblock Establishing the gzk cutoff with ultra high energy tau neutrinos.
\newblock In \emph{AIP Conference Proceedings}, pages 157--171, 2001.

\bibitem[{Feng} et~al.(2002){Feng}, {Fisher}, {Wilczek}, and
  {Yu}]{earth-skimming-feng}
J.~L. {Feng}, P.~{Fisher}, F.~{Wilczek}, and T.~M. {Yu}.
\newblock {Observability of Earth-Skimming Ultrahigh Energy Neutrinos}.
\newblock \emph{Physical Review Letters}, 88\penalty0 (16):\penalty0 161102,
  April 2002.

\bibitem[{Fargion} et~al.(2004){Fargion}, {De Sanctis Lucentini}, {De Santis},
  and {Grossi}]{fargion-3}
D.~{Fargion}, P.~G. {De Sanctis Lucentini}, M.~{De Santis}, and M.~{Grossi}.
\newblock {Tau Air Showers from Earth}.
\newblock \emph{Astrophysical Journal}, 613:\penalty0 1285--1301, October 2004.

\bibitem[{Conrad} et~al.(2010){Conrad}, {de Gouv{\^e}a}, {Shalgar}, and
  {Spitz}]{taudetections}
J.~{Conrad}, A.~{de Gouv{\^e}a}, S.~{Shalgar}, and J.~{Spitz}.
\newblock {Atmospheric tau neutrinos in a multikiloton liquid argon detector}.
\newblock \emph{Physical Review D}, 82\penalty0 (9):\penalty0 093012, November
  2010.

\bibitem[{Aartsen} et~al.(2016{\natexlab{a}}){Aartsen}, {Abraham}, {Ackermann},
  {Adams}, {Aguilar}, {Ahlers}, {Ahrens}, {Altmann}, {Anderson}, {Ansseau}, and
  et~al.]{icecube-taus}
M.~G. {Aartsen}, K.~{Abraham}, M.~{Ackermann}, J.~{Adams}, J.~A. {Aguilar},
  M.~{Ahlers}, M.~{Ahrens}, D.~{Altmann}, T.~{Anderson}, I.~{Ansseau}, and
  et~al.
\newblock {Search for astrophysical tau neutrinos in three years of IceCube
  data}.
\newblock \emph{Physical Review D}, 93\penalty0 (2):\penalty0 022001, January
  2016{\natexlab{a}}.

\bibitem[{Aab} et~al.(2015){Aab}, {Abreu}, {Aglietta}, {Ahn}, {Al Samarai},
  {Albuquerque}, {Allekotte}, {Allison}, {Almela}, {Alvarez Castillo}, and
  et~al.]{auger-es-3}
A.~{Aab}, P.~{Abreu}, M.~{Aglietta}, E.~J. {Ahn}, I.~{Al Samarai}, I.~F.~M.
  {Albuquerque}, I.~{Allekotte}, P.~{Allison}, A.~{Almela}, J.~{Alvarez
  Castillo}, and et~al.
\newblock {Improved limit to the diffuse flux of ultrahigh energy neutrinos
  from the Pierre Auger Observatory}.
\newblock \emph{Physical Review D}, 91\penalty0 (9):\penalty0 092008, May 2015.

\bibitem[{Abraham} et~al.(2008){Abraham}, {Abreu}, {Aglietta}, {Aguirre},
  {Allard}, {Allekotte}, {Allen}, {Allison}, {Alvarez-Mu{\~n}iz}, {Ambrosio},
  and et~al.]{auger-es-1}
J.~{Abraham}, P.~{Abreu}, M.~{Aglietta}, C.~{Aguirre}, D.~{Allard},
  I.~{Allekotte}, J.~{Allen}, P.~{Allison}, J.~{Alvarez-Mu{\~n}iz},
  M.~{Ambrosio}, and et~al.
\newblock {Upper Limit on the Diffuse Flux of Ultrahigh Energy Tau Neutrinos
  from the Pierre Auger Observatory}.
\newblock \emph{Physical Review Letters}, 100\penalty0 (21):\penalty0 211101,
  May 2008.

\bibitem[{Abraham} et~al.(2009){Abraham}, {Abreu}, {Aglietta}, {Aguirre},
  {Ahn}, {Allard}, {Allekotte}, {Allen}, {Allison}, {Alvarez-Mu{\~n}iz}, and
  et~al.]{auger-es-2}
J.~{Abraham}, P.~{Abreu}, M.~{Aglietta}, C.~{Aguirre}, E.~J. {Ahn},
  D.~{Allard}, I.~{Allekotte}, J.~{Allen}, P.~{Allison},
  J.~{Alvarez-Mu{\~n}iz}, and et~al.
\newblock {Limit on the diffuse flux of ultrahigh energy tau neutrinos with the
  surface detector of the Pierre Auger Observatory}.
\newblock \emph{Physical Review D}, 79\penalty0 (10):\penalty0 102001, May
  2009.

\bibitem[{Miele} et~al.(2006){Miele}, {Pastor}, and {Pisanti}]{auger-fluor}
G.~{Miele}, S.~{Pastor}, and O.~{Pisanti}.
\newblock {The aperture for UHE tau neutrinos of the Auger fluorescence
  detector using a Digital Elevation Map}.
\newblock \emph{Physics Letters B}, 634:\penalty0 137--142, March 2006.

\bibitem[{Aita} et~al.(2011){Aita}, {Aoki}, {Asaoka}, {Chonan}, {Jobashi},
  {Masuda}, {Morimoto}, {Noda}, {Sasaki}, {Asoh}, {Ishikawa}, {Ogawa},
  {Learned}, {Matsuno}, {Olsen}, {Binder}, {Hamilton}, {Sugiyama}, {Watanabe},
  and {Ashra-1 Collaboration}]{ashra-es-1}
Y.~{Aita}, T.~{Aoki}, Y.~{Asaoka}, T.~{Chonan}, M.~{Jobashi}, M.~{Masuda},
  Y.~{Morimoto}, K.~{Noda}, M.~{Sasaki}, J.~{Asoh}, N.~{Ishikawa}, S.~{Ogawa},
  J.~G. {Learned}, S.~{Matsuno}, S.~{Olsen}, P.-M. {Binder}, J.~{Hamilton},
  N.~{Sugiyama}, Y.~{Watanabe}, and {Ashra-1 Collaboration}.
\newblock {Observational Search for PeV-EeV Tau Neutrino from GRB081203A}.
\newblock \emph{Astrophysical Journal, Letters}, 736:\penalty0 L12, July 2011.

\bibitem[{Gaug} et~al.(2008){Gaug}, {Hsu}, {Becker}, {Biland}, {Mariotti},
  {Rhode}, and {Teshima}]{taus-magic-1}
M.~{Gaug}, C.~{Hsu}, J.~K. {Becker}, A.~{Biland}, M.~{Mariotti}, W.~{Rhode},
  and M.~{Teshima}.
\newblock {Tau neutrino search with the MAGIC telescope}.
\newblock volume~3, pages 1273--1276, 2008.

\bibitem[Becker et~al.(2008)Becker, Gaug, Hsu, Rhode, and
  Collaboration]{taus-magic-2}
Julia~K. Becker, Markus Gaug, Ching‐Cheng Hsu, Wolfgang Rhode, and MAGIC
  Collaboration.
\newblock Grb neutrino search with magic.
\newblock volume 1000, pages 245--248, 2008.

\bibitem[{Fargion} et~al.(2008{\natexlab{a}}){Fargion}, {Oliva}, {Massa}, and
  {Moreno}]{fargion-5}
D.~{Fargion}, P.~{Oliva}, F.~{Massa}, and G.~{Moreno}.
\newblock {Cherenkov flashes and fluorescence flares on telescopes: New lights
  on UHECR spectroscopy while unveiling neutrinos astronomy}.
\newblock \emph{Nuclear Instruments and Methods in Physics Research A},
  588:\penalty0 146--150, April 2008{\natexlab{a}}.

\bibitem[{Fargion} et~al.(2008{\natexlab{b}}){Fargion}, {Gaug}, and
  {Oliva}]{fargion-4}
D.~{Fargion}, M.~{Gaug}, and P.~{Oliva}.
\newblock {Reflecting on {\v C}erenkov reflections}.
\newblock In \emph{Journal of Physics Conference Series}, volume 110 of
  \emph{Journal of Physics Conference Series}, page 062008, May
  2008{\natexlab{b}}.

\bibitem[{Lipari} and {Stanev}(1991)]{stanev}
P.~{Lipari} and T.~{Stanev}.
\newblock {Propagation of multi-TeV muons}.
\newblock \emph{Physical Review D}, 44:\penalty0 3543--3554, December 1991.

\bibitem[{Smith for the HAWC Collaboration}(2015)]{AndyCal}
{Smith for the HAWC Collaboration}.
\newblock {HAWC: Design, Operation, Reconstruction and Analysis}.
\newblock August 2015.

\bibitem[{Abeysekara} et~al.(2017){Abeysekara}, {Albert}, {Alfaro}, {Alvarez},
  {{\'A}lvarez}, {Arceo}, {Arteaga-Vel{\'a}zquez}, {Ayala Solares}, {Barber},
  {Bautista-Elivar}, {Becerril}, {Belmont-Moreno}, {BenZvi}, {Berley}, {Braun},
  {Brisbois}, {Caballero-Mora}, {Capistr{\'a}n}, {Carrami{\~n}ana}, {Casanova},
  {Castillo}, {Cotti}, {Cotzomi}, {Couti{\~n}o de Le{\'o}n}, {de la Fuente},
  {De Le{\'o}n}, {DeYoung}, {Dingus}, {DuVernois}, {D{\'{\i}}az-V{\'e}lez},
  {Ellsworth}, {Fiorino}, {Fraija}, {Garc{\'{\i}}a-Gonz{\'a}lez}, {Gerhardt},
  {Gonz{\'a}lez Munoz}, {Gonz{\'a}lez}, {Goodman}, {Hampel-Arias}, {Harding},
  {Hernandez}, {Hernandez-Almada}, {Hinton}, {Hui}, {H{\"u}ntemeyer},
  {Iriarte}, {Jardin-Blicq}, {Joshi}, {Kaufmann}, {Kieda}, {Lara}, {Lauer},
  {Lee}, {Lennarz}, {Le{\'o}n Vargas}, {Linnemann}, {Longinotti}, {Raya},
  {Luna-Garc{\'{\i}}a}, {L{\'o}pez-Coto}, {Malone}, {Marinelli}, {Martinez},
  {Martinez-Castellanos}, {Mart{\'{\i}}nez-Castro}, {Mart{\'{\i}}nez-Huerta},
  {Matthews}, {Miranda-Romagnoli}, {Moreno}, {Mostaf{\'a}}, {Nellen},
  {Newbold}, {Nisa}, {Noriega-Papaqui}, {Pelayo}, {Pretz},
  {P{\'e}rez-P{\'e}rez}, {Ren}, {Rho}, {Rivi{\`e}re}, {Rosa-Gonz{\'a}lez},
  {Rosenberg}, {Ruiz-Velasco}, {Salazar}, {Salesa Greus}, {Sandoval},
  {Schneider}, {Schoorlemmer}, {Sinnis}, {Smith}, {Springer}, {Surajbali},
  {Taboada}, {Tibolla}, {Tollefson}, {Torres}, {Ukwatta}, {Villase{\~n}or},
  {Weisgarber}, {Westerhoff}, {Wisher}, {Wood}, {Yapici}, {Yodh}, {Younk},
  {Zepeda}, and {Zhou}]{Crab}
A.~U. {Abeysekara}, A.~{Albert}, R.~{Alfaro}, C.~{Alvarez}, J.~D.
  {{\'A}lvarez}, R.~{Arceo}, J.~C. {Arteaga-Vel{\'a}zquez}, H.~A. {Ayala
  Solares}, A.~S. {Barber}, N.~{Bautista-Elivar}, A.~{Becerril},
  E.~{Belmont-Moreno}, S.~Y. {BenZvi}, D.~{Berley}, J.~{Braun}, C.~{Brisbois},
  K.~S. {Caballero-Mora}, T.~{Capistr{\'a}n}, A.~{Carrami{\~n}ana},
  S.~{Casanova}, M.~{Castillo}, U.~{Cotti}, J.~{Cotzomi}, S.~{Couti{\~n}o de
  Le{\'o}n}, E.~{de la Fuente}, C.~{De Le{\'o}n}, T.~{DeYoung}, B.~L. {Dingus},
  M.~A. {DuVernois}, J.~C. {D{\'{\i}}az-V{\'e}lez}, R.~W. {Ellsworth}, D.~W.
  {Fiorino}, N.~{Fraija}, J.~A. {Garc{\'{\i}}a-Gonz{\'a}lez}, M.~{Gerhardt},
  A.~{Gonz{\'a}lez Munoz}, M.~M. {Gonz{\'a}lez}, J.~A. {Goodman},
  Z.~{Hampel-Arias}, J.~P. {Harding}, S.~{Hernandez}, A.~{Hernandez-Almada},
  J.~{Hinton}, C.~M. {Hui}, P.~{H{\"u}ntemeyer}, A.~{Iriarte},
  A.~{Jardin-Blicq}, V.~{Joshi}, S.~{Kaufmann}, D.~{Kieda}, A.~{Lara}, R.~J.
  {Lauer}, W.~H. {Lee}, D.~{Lennarz}, H.~{Le{\'o}n Vargas}, J.~T. {Linnemann},
  A.~L. {Longinotti}, G.~L. {Raya}, R.~{Luna-Garc{\'{\i}}a},
  R.~{L{\'o}pez-Coto}, K.~{Malone}, S.~S. {Marinelli}, O.~{Martinez},
  I.~{Martinez-Castellanos}, J.~{Mart{\'{\i}}nez-Castro},
  H.~{Mart{\'{\i}}nez-Huerta}, J.~A. {Matthews}, P.~{Miranda-Romagnoli},
  E.~{Moreno}, M.~{Mostaf{\'a}}, L.~{Nellen}, M.~{Newbold}, M.~U. {Nisa},
  R.~{Noriega-Papaqui}, R.~{Pelayo}, J.~{Pretz}, E.~G. {P{\'e}rez-P{\'e}rez},
  Z.~{Ren}, C.~D. {Rho}, C.~{Rivi{\`e}re}, D.~{Rosa-Gonz{\'a}lez},
  M.~{Rosenberg}, E.~{Ruiz-Velasco}, H.~{Salazar}, F.~{Salesa Greus},
  A.~{Sandoval}, M.~{Schneider}, H.~{Schoorlemmer}, G.~{Sinnis}, A.~J. {Smith},
  R.~W. {Springer}, P.~{Surajbali}, I.~{Taboada}, O.~{Tibolla}, K.~{Tollefson},
  I.~{Torres}, T.~N. {Ukwatta}, L.~{Villase{\~n}or}, T.~{Weisgarber},
  S.~{Westerhoff}, I.~G. {Wisher}, J.~{Wood}, T.~{Yapici}, G.~B. {Yodh}, P.~W.
  {Younk}, A.~{Zepeda}, and H.~{Zhou}.
\newblock {Observation of the Crab Nebula with the HAWC Gamma-Ray Observatory}.
\newblock \emph{ArXiv e-prints}, January 2017.

\bibitem[{Ayala Solares} et~al.(2015){Ayala Solares}, {Gerhardt}, {Hui},
  {Lauer}, {Ren}, {Salesa Greus}, {Zhou}, and {for the HAWC
  Collaboration}]{ICRCCal}
H.~A. {Ayala Solares}, M.~{Gerhardt}, C.~M. {Hui}, R.~J. {Lauer}, Z.~{Ren},
  F.~{Salesa Greus}, H.~{Zhou}, and {for the HAWC Collaboration}.
\newblock {The Calibration System of the HAWC Gamma-Ray Observatory}.
\newblock August 2015.

\bibitem[Agostinelli et~al.(2003)]{geant}
S.~Agostinelli et~al.
\newblock {GEANT4: A Simulation toolkit}.
\newblock \emph{Nucl. Instrum. Meth.}, A506:\penalty0 250--303, 2003.
\newblock \doi{10.1016/S0168-9002(03)01368-8}.

\bibitem[{Ata{\u g}} and {G{\"u}rkanl{\i}}(2016)]{colimation}
S.~{Ata{\u g}} and E.~{G{\"u}rkanl{\i}}.
\newblock {Prediction for CP violation via electric dipole moment of {$\tau$}
  lepton in {$\gamma$}{$\gamma$} $\rightarrow$ {$\tau$} $^{+}$ {$\tau$} $^{-}$
  process at CLIC}.
\newblock \emph{Journal of High Energy Physics}, 6:\penalty0 118, June 2016.

\bibitem[{Feng} et~al.(2001){Feng}, {Fisher}, {Wilczek}, and
  {Yu}]{earth-skimming-feng-pre}
J.~L. {Feng}, P.~{Fisher}, F.~{Wilczek}, and T.~M. {Yu}.
\newblock {Observability of Earth-Skimming Ultrahigh Energy Neutrinos}.
\newblock preprint, hep-ph/0105067v1, 2001.

\bibitem[{Dutta} et~al.(2005){Dutta}, {Huang}, and {Reno}]{tauloss3}
S.~I. {Dutta}, Y.~{Huang}, and M.~H. {Reno}.
\newblock {Tau neutrino propagation and tau energy loss}.
\newblock \emph{Physical Review D}, 72\penalty0 (1):\penalty0 013005, July
  2005.

\bibitem[{Blanch Bigas} et~al.(2008){Blanch Bigas}, {Deligny}, {Payet}, and
  {van Elewyck}]{tau-energy-loss-2}
O.~{Blanch Bigas}, O.~{Deligny}, K.~{Payet}, and V.~{van Elewyck}.
\newblock {Tau energy losses at ultrahigh energy: Continuous versus stochastic
  treatment}.
\newblock \emph{Physical Review D}, 77\penalty0 (10):\penalty0 103004, May
  2008.

\bibitem[{Iyer Dutta} et~al.(2001){Iyer Dutta}, {Reno}, {Sarcevic}, and
  {Seckel}]{tauloss4}
S.~{Iyer Dutta}, M.~H. {Reno}, I.~{Sarcevic}, and D.~{Seckel}.
\newblock {Propagation of muons and taus at high energies}.
\newblock \emph{Physical Review D}, 63\penalty0 (9):\penalty0 094020, May 2001.

\bibitem[{The IceCube Collaboration} et~al.(2015{\natexlab{a}}){The IceCube
  Collaboration}, {Aartsen}, {Abraham}, {Ackermann}, {Adams}, {Aguilar},
  {Ahlers}, {Ahrens}, {Altmann}, {Anderson}, and et~al.]{IceCubeFlux1}
{The IceCube Collaboration}, M.~G. {Aartsen}, K.~{Abraham}, M.~{Ackermann},
  J.~{Adams}, J.~A. {Aguilar}, M.~{Ahlers}, M.~{Ahrens}, D.~{Altmann},
  T.~{Anderson}, and et~al.
\newblock {The IceCube Neutrino Observatory - Contributions to ICRC 2015 Part
  II: Atmospheric and Astrophysical Diffuse Neutrino Searches of All Flavors:
  Combined Analysis of the High-Energy Cosmic Neutrino Flux at the IceCube
  Detector}.
\newblock October 2015{\natexlab{a}}.

\bibitem[{The IceCube Collaboration} et~al.(2015{\natexlab{b}}){The IceCube
  Collaboration}, {Aartsen}, {Abraham}, {Ackermann}, {Adams}, {Aguilar},
  {Ahlers}, {Ahrens}, {Altmann}, {Anderson}, and et~al.]{IceCubeFlux2}
{The IceCube Collaboration}, M.~G. {Aartsen}, K.~{Abraham}, M.~{Ackermann},
  J.~{Adams}, J.~A. {Aguilar}, M.~{Ahlers}, M.~{Ahrens}, D.~{Altmann},
  T.~{Anderson}, and et~al.
\newblock {The IceCube Neutrino Observatory - Contributions to ICRC 2015 Part
  II: Atmospheric and Astrophysical Diffuse Neutrino Searches of All Flavors: A
  measurement of the diffuse astrophysical muon neutrino flux using multiple
  years of IceCube data}.
\newblock October 2015{\natexlab{b}}.

\bibitem[{Aartsen} et~al.(2015){Aartsen}, {Abraham}, {Ackermann}, {Adams},
  {Aguilar}, {Ahlers}, {Ahrens}, {Altmann}, {Anderson}, {Archinger}, and
  et~al.]{IceCubeFlux3}
M.~G. {Aartsen}, K.~{Abraham}, M.~{Ackermann}, J.~{Adams}, J.~A. {Aguilar},
  M.~{Ahlers}, M.~{Ahrens}, D.~{Altmann}, T.~{Anderson}, M.~{Archinger}, and
  et~al.
\newblock {A Combined Maximum-likelihood Analysis of the High-energy
  Astrophysical Neutrino Flux Measured with IceCube}.
\newblock \emph{Astrophysical Journal}, 809:\penalty0 98, August 2015.

\bibitem[{Schoenen} and {Raedel}(2015)]{UHENeutIC}
S.~{Schoenen} and L.~{Raedel}.
\newblock {Detection of a multi-PeV neutrino-induced muon event from the
  Northern sky with IceCube}.
\newblock \emph{The Astronomer's Telegram}, 7856, July 2015.

\bibitem[{Kistler} and {Laha}(2016)]{UHENMod}
M.~D. {Kistler} and R.~{Laha}.
\newblock {Multi-PeV Signals from a New Astrophysical Neutrino Flux Beyond the
  Glashow Resonance}.
\newblock May 2016.

\bibitem[{The IceCube Collaboration} et~al.(2015{\natexlab{c}}){The IceCube
  Collaboration}, {Aartsen}, {Abraham}, {Ackermann}, {Adams}, {Aguilar},
  {Ahlers}, {Ahrens}, {Altmann}, {Anderson}, and et~al.]{IceCubeFlux4}
{The IceCube Collaboration}, M.~G. {Aartsen}, K.~{Abraham}, M.~{Ackermann},
  J.~{Adams}, J.~A. {Aguilar}, M.~{Ahlers}, M.~{Ahrens}, D.~{Altmann},
  T.~{Anderson}, and et~al.
\newblock {The IceCube Neutrino Observatory - Contributions to ICRC 2015 Part
  II: Atmospheric and Astrophysical Diffuse Neutrino Searches of All Flavors:
  Observation of Astrophysical Neutrinos in Four Years of IceCube Data}.
\newblock October 2015{\natexlab{c}}.

\bibitem[{Armesto} et~al.(2008){Armesto}, {Merino}, {Parente}, and
  {Zas}]{tau-energy-loss}
N.~{Armesto}, C.~{Merino}, G.~{Parente}, and E.~{Zas}.
\newblock {Charged current neutrino cross section and tau energy loss at
  ultrahigh energies}.
\newblock \emph{Physical Review D}, 77\penalty0 (1):\penalty0 013001, January
  2008.

\bibitem[{Gandhi} et~al.(1998){Gandhi}, {Quigg}, {Reno}, and
  {Sarcevic}]{gandhi}
R.~{Gandhi}, C.~{Quigg}, M.~H. {Reno}, and I.~{Sarcevic}.
\newblock {Neutrino interactions at ultrahigh energies}.
\newblock \emph{Physical Review D}, 58\penalty0 (9):\penalty0 093009, November
  1998.

\bibitem[Connolly et~al.(2011)Connolly, Thorne, and Waters]{conolly}
Amy Connolly, Robert~S. Thorne, and David Waters.
\newblock Calculation of high energy neutrino-nucleon cross sections and
  uncertainties using the martin-stirling-thorne-watt parton distribution
  functions and implications for future experiments.
\newblock \emph{Physical Review D}, 83:\penalty0 113009, Jun 2011.

\bibitem[{Arg{\"u}elles} et~al.(2015){Arg{\"u}elles}, {Halzen}, {Wille},
  {Kroll}, and {Reno}]{arguelles}
C.~A. {Arg{\"u}elles}, F.~{Halzen}, L.~{Wille}, M.~{Kroll}, and M.~H. {Reno}.
\newblock {High-energy behavior of photon, neutrino, and proton cross
  sections}.
\newblock \emph{Physical Review D}, 92\penalty0 (7):\penalty0 074040, October
  2015.

\bibitem[Cho(2013)]{science-letter-neutrinos}
Adrian Cho.
\newblock Physicists snare a precious few neutrinos from the cosmos.
\newblock \emph{Science}, 342\penalty0 (6161):\penalty0 920--920, 2013.
\newblock ISSN 0036-8075.

\bibitem[{Abeysekara} et~al.(2014){Abeysekara}, {Alfaro}, {Alvarez},
  {{\'A}lvarez}, {Arceo}, {Arteaga-Vel{\'a}zquez}, {Ayala Solares}, {Barber},
  {Baughman}, {Bautista-Elivar}, {Belmont}, {BenZvi}, {Berley}, {Bonilla
  Rosales}, {Braun}, {Caballero-Mora}, {Carrami{\~n}ana}, {Castillo}, {Cotti},
  {Cotzomi}, {de la Fuente}, {De Le{\'o}n}, {DeYoung}, {Diaz Hernandez},
  {D{\'{\i}}az-V{\'e}lez}, {Dingus}, {DuVernois}, {Ellsworth}, {Fiorino},
  {Fraija}, {Galindo}, {Garfias}, {Gonz{\'a}lez}, {Goodman}, {Gussert},
  {Hampel-Arias}, {Harding}, {H{\"u}ntemeyer}, {Hui}, {Imran}, {Iriarte},
  {Karn}, {Kieda}, {Kunde}, {Lara}, {Lauer}, {Lee}, {Lennarz}, {Le{\'o}n
  Vargas}, {Linnemann}, {Longo}, {Luna-Garc{\'{\i}}a}, {Malone}, {Marinelli},
  {Marinelli}, {Martinez}, {Martinez}, {Mart{\'{\i}}nez-Castro}, {Matthews},
  {McEnery}, {Mendoza Torres}, {Miranda-Romagnoli}, {Moreno}, {Mostaf{\'a}},
  {Nellen}, {Newbold}, {Noriega-Papaqui}, {Oceguera-Becerra}, {Patricelli},
  {Pelayo}, {P{\'e}rez-P{\'e}rez}, {Pretz}, {Rivi{\`e}re}, {Rosa-Gonz{\'a}lez},
  {Ruiz-Velasco}, {Ryan}, {Salazar}, {Salesa Greus}, {Sandoval}, {Schneider},
  {Sinnis}, {Smith}, {Sparks Woodle}, {Springer}, {Taboada}, {Toale},
  {Tollefson}, {Torres}, {Ukwatta}, {Villase{\~n}or}, {Weisgarber},
  {Westerhoff}, {Wisher}, {Wood}, {Yodh}, {Younk}, {Zaborov}, {Zepeda}, {Zhou},
  and {HAWC Collaboration}]{hawc-ani}
A.~U. {Abeysekara}, R.~{Alfaro}, C.~{Alvarez}, J.~D. {{\'A}lvarez}, R.~{Arceo},
  J.~C. {Arteaga-Vel{\'a}zquez}, H.~A. {Ayala Solares}, A.~S. {Barber}, B.~M.
  {Baughman}, N.~{Bautista-Elivar}, E.~{Belmont}, S.~Y. {BenZvi}, D.~{Berley},
  M.~{Bonilla Rosales}, J.~{Braun}, K.~S. {Caballero-Mora},
  A.~{Carrami{\~n}ana}, M.~{Castillo}, U.~{Cotti}, J.~{Cotzomi}, E.~{de la
  Fuente}, C.~{De Le{\'o}n}, T.~{DeYoung}, R.~{Diaz Hernandez}, J.~C.
  {D{\'{\i}}az-V{\'e}lez}, B.~L. {Dingus}, M.~A. {DuVernois}, R.~W.
  {Ellsworth}, D.~W. {Fiorino}, N.~{Fraija}, A.~{Galindo}, F.~{Garfias}, M.~M.
  {Gonz{\'a}lez}, J.~A. {Goodman}, M.~{Gussert}, Z.~{Hampel-Arias}, J.~P.
  {Harding}, P.~{H{\"u}ntemeyer}, C.~M. {Hui}, A.~{Imran}, A.~{Iriarte},
  P.~{Karn}, D.~{Kieda}, G.~J. {Kunde}, A.~{Lara}, R.~J. {Lauer}, W.~H. {Lee},
  D.~{Lennarz}, H.~{Le{\'o}n Vargas}, J.~T. {Linnemann}, M.~{Longo},
  R.~{Luna-Garc{\'{\i}}a}, K.~{Malone}, A.~{Marinelli}, S.~S. {Marinelli},
  H.~{Martinez}, O.~{Martinez}, J.~{Mart{\'{\i}}nez-Castro}, J.~A.~J.
  {Matthews}, J.~{McEnery}, E.~{Mendoza Torres}, P.~{Miranda-Romagnoli},
  E.~{Moreno}, M.~{Mostaf{\'a}}, L.~{Nellen}, M.~{Newbold},
  R.~{Noriega-Papaqui}, T.~{Oceguera-Becerra}, B.~{Patricelli}, R.~{Pelayo},
  E.~G. {P{\'e}rez-P{\'e}rez}, J.~{Pretz}, C.~{Rivi{\`e}re},
  D.~{Rosa-Gonz{\'a}lez}, E.~{Ruiz-Velasco}, J.~{Ryan}, H.~{Salazar},
  F.~{Salesa Greus}, A.~{Sandoval}, M.~{Schneider}, G.~{Sinnis}, A.~J. {Smith},
  K.~{Sparks Woodle}, R.~W. {Springer}, I.~{Taboada}, P.~A. {Toale},
  K.~{Tollefson}, I.~{Torres}, T.~N. {Ukwatta}, L.~{Villase{\~n}or},
  T.~{Weisgarber}, S.~{Westerhoff}, I.~G. {Wisher}, J.~{Wood}, G.~B. {Yodh},
  P.~W. {Younk}, D.~{Zaborov}, A.~{Zepeda}, H.~{Zhou}, and {HAWC
  Collaboration}.
\newblock {Observation of Small-scale Anisotropy in the Arrival Direction
  Distribution of TeV Cosmic Rays with HAWC}.
\newblock \emph{Astrophysical Journal}, 796:\penalty0 108, December 2014.

\bibitem[{Kusenko} and {Weiler}(2002)]{earth-skimming-prl2}
A.~{Kusenko} and T.~J. {Weiler}.
\newblock {Neutrino Cross Sections and Future Observations of Ultrahigh-Energy
  Cosmic Rays}.
\newblock \emph{Physical Review Letters}, 88\penalty0 (16):\penalty0 161101,
  April 2002.

\bibitem[{G{\'o}ra} et~al.(2015){G{\'o}ra}, {Bernardini}, and
  {Kappes}]{tau-neutrinos-mc1}
D.~{G{\'o}ra}, E.~{Bernardini}, and A.~{Kappes}.
\newblock {Searching for tau neutrinos with Cherenkov telescopes}.
\newblock \emph{Astroparticle Physics}, 61:\penalty0 12--16, February 2015.

\bibitem[Cao et~al.(2005)Cao, Huang, Sokolsky, and Hu]{eff-area-taus-fullmc}
Z~Cao, M~A Huang, P~Sokolsky, and Y~Hu.
\newblock Ultra high energy ν τ detection with a cosmic ray tau neutrino
  telescope using fluorescence/cerenkov light technique.
\newblock \emph{Journal of Physics G: Nuclear and Particle Physics},
  31\penalty0 (7):\penalty0 571, 2005.

\bibitem[INEGI(2012)]{inegi}
INEGI.
\newblock Continuo de elevaciones mexicano 3.0 (cem 3.0), 2012.
\newblock URL
  \url{http://www.inegi.org.mx/geo/contenidos/datosrelieve/continental/continuoelevaciones.aspx}.
\newblock Cited 25 Dec 1999.

\bibitem[{Aartsen} et~al.(2016{\natexlab{b}}){Aartsen}, {Abraham}, {Ackermann},
  {Adams}, {Aguilar}, {Ahlers}, {Ahrens}, {Altmann}, {Anderson}, {Archinger},
  and et~al.]{atmos-mu-ic}
M.~G. {Aartsen}, K.~{Abraham}, M.~{Ackermann}, J.~{Adams}, J.~A. {Aguilar},
  M.~{Ahlers}, M.~{Ahrens}, D.~{Altmann}, T.~{Anderson}, M.~{Archinger}, and
  et~al.
\newblock {Characterization of the atmospheric muon flux in IceCube}.
\newblock \emph{Astroparticle Physics}, 78:\penalty0 1--27, May
  2016{\natexlab{b}}.

\bibitem[{Anchordoqui} et~al.(2014){Anchordoqui}, {Barger}, {Cholis},
  {Goldberg}, {Hooper}, {Kusenko}, {Learned}, {Marfatia}, {Pakvasa}, {Paul},
  and {Weiler}]{atmos-mu-2}
L.~A. {Anchordoqui}, V.~{Barger}, I.~{Cholis}, H.~{Goldberg}, D.~{Hooper},
  A.~{Kusenko}, J.~G. {Learned}, D.~{Marfatia}, S.~{Pakvasa}, T.~C. {Paul}, and
  T.~J. {Weiler}.
\newblock {Cosmic neutrino pevatrons: A brand new pathway to astronomy,
  astrophysics, and particle physics}.
\newblock \emph{Journal of High Energy Astrophysics}, 1:\penalty0 1--30, May
  2014.

\bibitem[{Abdo} et~al.(2015){Abdo}, {Abeysekara}, {Alfaro}, {Allen}, {Alvarez},
  {{\'A}lvarez}, {Arceo}, {Arteaga-Vel{\'a}zquez}, {Aune}, {Ayala Solares},
  {Barber}, {Baughman}, {Bautista-Elivar}, {Becerra Gonzalez}, {Belmont},
  {BenZvi}, {Berley}, {Bonilla Rosales}, {Braun}, {Caballero-Lopez},
  {Caballero-Mora}, {Carrami{\~n}ana}, {Castillo}, {Christopher}, {Cotti},
  {Cotzomi}, {de la Fuente}, {De Le{\'o}n}, {DeYoung}, {Diaz Hernandez},
  {Diaz-Cruz}, {D{\'{\i}}az-V{\'e}lez}, {Dingus}, {DuVernois}, {Ellsworth},
  {Fiorino}, {Fraija}, {Galindo}, {Garfias}, {Gonz{\'a}lez}, {Goodman},
  {Grabski}, {Gussert}, {Hampel-Arias}, {Harding}, {Hays}, {Hoffman}, {Hui},
  {H{\"u}ntemeyer}, {Imran}, {Iriarte}, {Karn}, {Kieda}, {Kolterman}, {Kunde},
  {Lara}, {Lauer}, {Lee}, {Lennarz}, {Le{\'o}n Vargas}, {Linares}, {Linnemann},
  {Longo}, {Luna-GarcIa}, {MacGibbon}, {Marinelli}, {Marinelli}, {Martinez},
  {Martinez}, {Mart{\'{\i}}nez-Castro}, {Matthews}, {McEnery}, {Mendoza
  Torres}, {Mincer}, {Miranda-Romagnoli}, {Moreno}, {Morgan}, {Mostaf{\'a}},
  {Nellen}, {Nemethy}, {Newbold}, {Noriega-Papaqui}, {Oceguera-Becerra},
  {Patricelli}, {Pelayo}, {P{\'e}rez-P{\'e}rez}, {Pretz}, {Rivi{\`e}re},
  {Rosa-Gonz{\'a}lez}, {Ruiz-Velasco}, {Ryan}, {Salazar}, {Salesa}, {Sandoval},
  {Saz Parkinson}, {Schneider}, {Silich}, {Sinnis}, {Smith}, {Stump}, {Sparks
  Woodle}, {Springer}, {Taboada}, {Toale}, {Tollefson}, {Torres}, {Ukwatta},
  {Vasileiou}, {Villase{\~n}or}, {Weisgarber}, {Westerhoff}, {Williams},
  {Wisher}, {Wood}, {Yodh}, {Younk}, {Zaborov}, {Zepeda}, and {Zhou}]{hawc-5}
A.~A. {Abdo}, A.~U. {Abeysekara}, R.~{Alfaro}, B.~T. {Allen}, C.~{Alvarez},
  J.~D. {{\'A}lvarez}, R.~{Arceo}, J.~C. {Arteaga-Vel{\'a}zquez}, T.~{Aune},
  H.~A. {Ayala Solares}, A.~S. {Barber}, B.~M. {Baughman},
  N.~{Bautista-Elivar}, J.~{Becerra Gonzalez}, E.~{Belmont}, S.~Y. {BenZvi},
  D.~{Berley}, M.~{Bonilla Rosales}, J.~{Braun}, R.~A. {Caballero-Lopez}, K.~S.
  {Caballero-Mora}, A.~{Carrami{\~n}ana}, M.~{Castillo}, G.~E. {Christopher},
  U.~{Cotti}, J.~{Cotzomi}, E.~{de la Fuente}, C.~{De Le{\'o}n}, T.~{DeYoung},
  R.~{Diaz Hernandez}, L.~{Diaz-Cruz}, J.~C. {D{\'{\i}}az-V{\'e}lez}, B.~L.
  {Dingus}, M.~A. {DuVernois}, R.~W. {Ellsworth}, D.~W. {Fiorino}, N.~{Fraija},
  A.~{Galindo}, F.~{Garfias}, M.~M. {Gonz{\'a}lez}, J.~A. {Goodman},
  V.~{Grabski}, M.~{Gussert}, Z.~{Hampel-Arias}, J.~P. {Harding}, E.~{Hays},
  C.~M. {Hoffman}, C.~M. {Hui}, P.~{H{\"u}ntemeyer}, A.~{Imran}, A.~{Iriarte},
  P.~{Karn}, D.~{Kieda}, B.~E. {Kolterman}, G.~J. {Kunde}, A.~{Lara}, R.~J.
  {Lauer}, W.~H. {Lee}, D.~{Lennarz}, H.~{Le{\'o}n Vargas}, E.~C. {Linares},
  J.~T. {Linnemann}, M.~{Longo}, R.~{Luna-GarcIa}, J.~H. {MacGibbon},
  A.~{Marinelli}, S.~S. {Marinelli}, H.~{Martinez}, O.~{Martinez},
  J.~{Mart{\'{\i}}nez-Castro}, J.~A.~J. {Matthews}, J.~{McEnery}, E.~{Mendoza
  Torres}, A.~I. {Mincer}, P.~{Miranda-Romagnoli}, E.~{Moreno}, T.~{Morgan},
  M.~{Mostaf{\'a}}, L.~{Nellen}, P.~{Nemethy}, M.~{Newbold},
  R.~{Noriega-Papaqui}, T.~{Oceguera-Becerra}, B.~{Patricelli}, R.~{Pelayo},
  E.~G. {P{\'e}rez-P{\'e}rez}, J.~{Pretz}, C.~{Rivi{\`e}re},
  D.~{Rosa-Gonz{\'a}lez}, E.~{Ruiz-Velasco}, J.~{Ryan}, H.~{Salazar},
  F.~{Salesa}, A.~{Sandoval}, P.~M. {Saz Parkinson}, M.~{Schneider},
  S.~{Silich}, G.~{Sinnis}, A.~J. {Smith}, D.~{Stump}, K.~{Sparks Woodle},
  R.~W. {Springer}, I.~{Taboada}, P.~A. {Toale}, K.~{Tollefson}, I.~{Torres},
  T.~N. {Ukwatta}, V.~{Vasileiou}, L.~{Villase{\~n}or}, T.~{Weisgarber},
  S.~{Westerhoff}, D.~A. {Williams}, I.~G. {Wisher}, J.~{Wood}, G.~B. {Yodh},
  P.~W. {Younk}, D.~{Zaborov}, A.~{Zepeda}, and H.~{Zhou}.
\newblock {Milagro limits and HAWC sensitivity for the rate-density of
  evaporating Primordial Black Holes}.
\newblock \emph{Astroparticle Physics}, 64:\penalty0 4--12, April 2015.

\bibitem[{Abeysekara} et~al.(2015){Abeysekara}, {Alfaro}, {Alvarez},
  {{\'A}lvarez}, {Arceo}, {Arteaga-Vel{\'a}zquez}, {Ayala Solares}, {Barber},
  {Baughman}, {Bautista-Elivar}, {BenZvi}, {Bonilla Rosales}, {Braun},
  {Caballero-Mora}, {Carrami{\~n}ana}, {Castillo}, {Cotti}, {Cotzomi}, {de la
  Fuente}, {De Le{\'o}n}, {DeYoung}, {Diaz Hernandez}, {Dingus}, {DuVernois},
  {Ellsworth}, {Fiorino}, {Fraija}, {Galindo}, {Garfias}, {Gonz{\'a}lez},
  {Goodman}, {Gussert}, {Hampel-Arias}, {Harding}, {H{\"u}ntemeyer}, {Hui},
  {Imran}, {Iriarte}, {Karn}, {Kieda}, {Kunde}, {Lara}, {Lauer}, {Lee},
  {Lennarz}, {Le{\'o}n Vargas}, {Linnemann}, {Longo}, {Luna-Garc{\'{\i}}a},
  {Malone}, {Marinelli}, {Marinelli}, {Martinez}, {Martinez},
  {Mart{\'{\i}}nez-Castro}, {Matthews}, {Mendoza Torres}, {Miranda-Romagnoli},
  {Moreno}, {Mostaf{\'a}}, {Nellen}, {Newbold}, {Noriega-Papaqui},
  {Oceguera-Becerra}, {Patricelli}, {Pelayo}, {P{\'e}rez-P{\'e}rez}, {Pretz},
  {Rivi{\`e}re}, {Rosa-Gonz{\'a}lez}, {Salazar}, {Salesa Greus}, {Sandoval},
  {Schneider}, {Sinnis}, {Smith}, {Sparks Woodle}, {Springer}, {Taboada},
  {Tollefson}, {Torres}, {Ukwatta}, {Villase{\~n}or}, {Weisgarber},
  {Westerhoff}, {Wisher}, {Wood}, {Yodh}, {Younk}, {Zaborov}, {Zepeda}, {Zhou},
  and {HAWC Collaboration}]{hawc-grbs}
A.~U. {Abeysekara}, R.~{Alfaro}, C.~{Alvarez}, J.~D. {{\'A}lvarez}, R.~{Arceo},
  J.~C. {Arteaga-Vel{\'a}zquez}, H.~A. {Ayala Solares}, A.~S. {Barber}, B.~M.
  {Baughman}, N.~{Bautista-Elivar}, S.~Y. {BenZvi}, M.~{Bonilla Rosales},
  J.~{Braun}, K.~S. {Caballero-Mora}, A.~{Carrami{\~n}ana}, M.~{Castillo},
  U.~{Cotti}, J.~{Cotzomi}, E.~{de la Fuente}, C.~{De Le{\'o}n}, T.~{DeYoung},
  R.~{Diaz Hernandez}, B.~L. {Dingus}, M.~A. {DuVernois}, R.~W. {Ellsworth},
  D.~W. {Fiorino}, N.~{Fraija}, A.~{Galindo}, F.~{Garfias}, M.~M.
  {Gonz{\'a}lez}, J.~A. {Goodman}, M.~{Gussert}, Z.~{Hampel-Arias}, J.~P.
  {Harding}, P.~{H{\"u}ntemeyer}, C.~M. {Hui}, A.~{Imran}, A.~{Iriarte},
  P.~{Karn}, D.~{Kieda}, G.~J. {Kunde}, A.~{Lara}, R.~J. {Lauer}, W.~H. {Lee},
  D.~{Lennarz}, H.~{Le{\'o}n Vargas}, J.~T. {Linnemann}, M.~{Longo},
  R.~{Luna-Garc{\'{\i}}a}, K.~{Malone}, A.~{Marinelli}, S.~S. {Marinelli},
  H.~{Martinez}, O.~{Martinez}, J.~{Mart{\'{\i}}nez-Castro}, J.~A.~J.
  {Matthews}, E.~{Mendoza Torres}, P.~{Miranda-Romagnoli}, E.~{Moreno},
  M.~{Mostaf{\'a}}, L.~{Nellen}, M.~{Newbold}, R.~{Noriega-Papaqui}, T.~O.
  {Oceguera-Becerra}, B.~{Patricelli}, R.~{Pelayo}, E.~G.
  {P{\'e}rez-P{\'e}rez}, J.~{Pretz}, C.~{Rivi{\`e}re}, D.~{Rosa-Gonz{\'a}lez},
  H.~{Salazar}, F.~{Salesa Greus}, A.~{Sandoval}, M.~{Schneider}, G.~{Sinnis},
  A.~J. {Smith}, K.~{Sparks Woodle}, R.~W. {Springer}, I.~{Taboada},
  K.~{Tollefson}, I.~{Torres}, T.~N. {Ukwatta}, L.~{Villase{\~n}or},
  T.~{Weisgarber}, S.~{Westerhoff}, I.~G. {Wisher}, J.~{Wood}, G.~B. {Yodh},
  P.~W. {Younk}, D.~{Zaborov}, A.~{Zepeda}, H.~{Zhou}, and {HAWC
  Collaboration}.
\newblock {Search for Gamma-Rays from the Unusually Bright GRB 130427A with the
  HAWC Gamma-Ray Observatory}.
\newblock \emph{Astrophysical Journal}, 800:\penalty0 78, February 2015.

\bibitem[{R{\"a}del} and {Wiebusch}(2013)]{CLight}
L.~{R{\"a}del} and C.~{Wiebusch}.
\newblock {Calculation of the Cherenkov light yield from electromagnetic
  cascades in ice with Geant4}.
\newblock \emph{Astroparticle Physics}, 44:\penalty0 102--113, April 2013.

\bibitem[Patrignani et~al.(2016)]{pdg}
C.~Patrignani et~al.
\newblock {Review of Particle Physics}.
\newblock \emph{Chin. Phys.}, C40\penalty0 (10):\penalty0 100001, 2016.

\bibitem[Jokisch et~al.(1979)Jokisch, Carstensen, Dau, Meyer, and
  Allkofer]{muonshz}
H.~Jokisch, K.~Carstensen, W.~D. Dau, H.~J. Meyer, and O.~C. Allkofer.
\newblock Cosmic-ray muon spectrum up to 1 tev at 75$^\circ$ zenith angle.
\newblock \emph{Phys. Rev. D}, 19:\penalty0 1368--1372, Mar 1979.

\bibitem[Sarcevic(2007)]{NXs-Sarcevic}
Ina Sarcevic.
\newblock Ultrahigh energy cosmic neutrinos and the physics beyond the standard
  model.
\newblock volume~60, page 175, 2007.

\end{thebibliography}

\end{document}